\documentclass[%
twocolumn,
nofootinbib,
nobibnotes,
 amsmath,amssymb,
 aps,prd,
floatfix,
]{revtex4-2}

\usepackage{graphicx}
\usepackage{dcolumn}
\usepackage{bm}
\usepackage{hyperref}
\usepackage[mathlines]{lineno}
\usepackage{graphicx}
\usepackage{dcolumn}
\usepackage{bm}
\usepackage{hyperref}
\usepackage{color}
\usepackage{natbib}

\usepackage{CJK}
\usepackage[normalem]{ulem}
\usepackage{comment}

\usepackage[english]{babel}
\usepackage{xspace}
\usepackage{ulem}

\hypersetup{
    colorlinks=true,
    linkcolor=magenta,
    citecolor=blue,
}

\newcommand{\POLUU}{\texttt{POL-UU}{}}
\newcommand{\HRPOLUU}{\texttt{HR-POL-UU}{}}
\newcommand{\POLUD}{\texttt{POL-UD}{}}
\newcommand{\TORPP}{\texttt{TOR-PP}{}}
\newcommand{\TORPM}{\texttt{TOR-PM}{}}
\newcommand{\HRMIX}{\texttt{HR-MIX}{}}
\newcommand{\MIX}{\texttt{MIX}{}}

\newcommand{\NOB}{\texttt{B0}{}}
\newcommand{\LOWB}{\texttt{LOWB}{}}
\newcommand{\POLLR}{\texttt{POL-90D}{}}
\newcommand{\POLASYM}{\texttt{POL-ASYM}{}}
\newcommand{\STIFF}{\texttt{STIFF}{}}

\newcommand{\BITANT}{\texttt{BITANT}{}}
\newcommand{\POLQ}{\texttt{POL-Q12}{}}

\begin{document}

\title{Magnetic Field Configurations in Binary Neutron Star Mergers I:\\
Post-merger Remnant and Disk}
\author{Eduardo M. Guti\'errez$^{1,2}$}\email{emgutierrez@psu.edu}
\author{William Cook$^3$}
\author{David Radice$^{1,2,4}$}
\author{Sebastiano Bernuzzi$^3$}
\author{Jacob Fields$^{1,2}$}
\author{Peter Hammond$^{1,2,5}$}
\author{Boris Daszuta$^{3}$}
\author{Harshraj Bandyopadhyay$^2$}
\author{Maximilian Jacobi$^3$}

\affiliation{%
$^1$Institute for Gravitation and the Cosmos, The Pennsylvania State University, University Park, Pennsylvania, 16802, USA \\
$^2$Department of Physics, The Pennsylvania State University, University Park, Pennsylvania, 16802, USA \\
$^3$Theoretisch-Physikalisches Institut, Friedrich-Schiller-Universit¨at Jena, 07743, Jena, Germany \\
$^4$Department of Astronomy and Astrophysics, The Pennsylvania State University, University Park, Pennsylvania, 16802, USA \\
$^5$Department of Physics and Astronomy, University of New Hampshire, Durham, NH 03824, USA
}
\date{June 2025}

\begin{abstract}
We present a suite of general relativistic magnetohydrodynamic (GRMHD) simulations of binary neutron star (BNS) mergers performed with the code {\tt GR-Athena++}.
We investigate how a different initial magnetic field configuration, nuclear equation of state, or binary mass ratio affects the magnetic and thermodynamic evolution of the post-merger remnant and disk.
We also analyze the impact of the commonly-assumed reflection (bitant) symmetry across the equatorial plane.
Magnetic field amplification occurs shortly after the merger due to the Kelvin--Helmholtz instability; later, the field keeps evolving with a predominantly toroidal configuration due to winding and turbulence.
The initial magnetic field topology leaves an imprint on the field structure and affects magnetic field amplification for the initial magnetic field values commonly assumed in the literature and the limited resolution of the simulations.
Enforcing equatorial reflection symmetry partially suppresses the development of turbulence near the equatorial plane and impacts the post-merger magnetic field evolution.
Stiffer EOSs produce larger, less compact remnants that may retain memory of the pre-merger strong poloidal field.
\end{abstract}

\maketitle

\section{Introduction}

Binary neutron star (BNS) mergers are among the most extreme phenomena in the Universe and provide a unique laboratory to study fundamental physics, including the interplay between extreme magnetic fields, gravitational fields, and the properties of dense nuclear matter.
The multimessenger detection of gravitational waves (GWs) and electromagnetic radiation (EM) from GW170817 \cite{LIGO_2017ApJ...848L..12A} marked a breakthrough in astronomy, confirming several long-standing hypotheses about phenomena associated to BNS mergers, such as the fact that the outflows from these events are key sites for the production of heavy elements via the rapid neutron-capture process ($r$-process) \cite{Kasen_etal2017Natur.551...80K, Cowperthwaite2017ApJ...848L..17C}, or that they can give rise to short $\gamma$-ray bursts (sGRBs) \cite{Eichler_etal1989Natur.340..126E, Goldstein_etal2017ApJ...848L..14G, LIGO_2017ApJ...848L..13A, Savchenko_etal2017ApJ...848L..15S}.
Recently, it has been proposed that BNS mergers can also produce long $\gamma$-ray bursts \cite{Gottlieb_etal2023ApJ...958L..33G, Gottlieb_etal2025ApJ...984...77G, Perna_etal2025PhRvD.111f3015P}.

One important factor influencing the EM counterparts of BNS mergers is the structure and evolution of magnetic fields, since they may shape the post-merger dynamics and drive the launching of outflows.
Neutron stars in binary systems have surface magnetic fields of ${\sim} 10^8-10^{12}~{\rm G}$ \cite{Lorimer2008LRR....11....8L}, which may be amplified during and after the merger up to values of up to ${\sim} 10^{16}~{\rm G}$ \cite{Kiuchi_etal2015PhRvD..92l4034K, Kiuchi_etal2018PhRvD..97l4039K, Kiuchi_etal2024NatAs...8..298K, Giacomazzo_etal2015ApJ...809...39G, Rezzolla_etal2011ApJ...732L...6R, CombiSiegel2023PhRvL.131w1402C, Palenzuela_etal2022PhRvD.106b3013P, AguileraMiret_etal2022ApJ...926L..31A, AguileraMiret_etal2023PhRvD.108j3001A, AguileraMiret_etal2024PhRvD.110h3014A, AguileraMiret_etal2025arXiv250410604A}.
Such amplification is triggered by the development of magnetohydrodynamic (MHD) instabilities, such as the Kelvin--Helmholtz (KH) instability in the shear layer between the colliding stars \cite{RasioShapiro1999CQGra..16R...1R, Obergaulinger_etal2006A&A...450.1107O, Kiuchi_etal2014PhRvD..90d1502K, Kiuchi_etal2015PhRvD..92l4034K, ZrakeMacFadyen2013ApJ...769L..29Z}, and the magnetorotational instability (MRI) in the outer layers of the post-merger remnant and throughout the surrounding disk \cite{BalbusHawley1991ApJ...376..214B, CombiSiegel2023PhRvL.131w1402C, Kiuchi_etal2024NatAs...8..298K, Celora_etal2025arXiv250501208C}.
Subsequently, additional mechanisms such as magnetic winding \cite{Duez_etal2006PhRvL..96c1101D, Shibata_etal2021PhRvD.103d3022S, AguileraMiret_etal2023PhRvD.108j3001A} and the Tayler--Spruit dynamo \cite{Spruit2002A&A...381..923S, ReboulSalze_etal2025A&A...699A...4R} may further amplify the magnetic field. The latter process, however, also transports angular momentum and may reduce the differential rotation and saturate the toroidal field.
The field could rise to the surface of the remnant through the Parker instability \cite{MostQuataert2023ApJ...947L..15M, Jiang_etal2025PhRvD.111j3043J} and power strong flares, although it remains uncertain whether this occurs unless the field strength becomes excessively large \cite{Fields_etal2025arXiv250718695F}.
The highly magnetized post-merger remnant may also serve as an engine to launch powerful outflows which can collimate into relativistic jets \cite{Rezzolla_etal2011ApJ...732L...6R, Ciolfi2020MNRAS.495L..66C, Moesta_etal2020ApJ...901L..37M, CombiSiegel2023PhRvL.131w1402C, Kiuchi_etal2023PhRvL.131a1401K, Ruiz_etal2016ApJ...824L...6R, Ruiz_etal2020PhRvD.101f4042R, Ruiz_etal2021PhRvD.104l4049R, Bamber_etal2024PhRvD.110b4046B} and thus power sGRBs.

Several key aspects of the magnetic field evolution in BNS mergers remain uncertain, including the precise mechanisms and timescales for its amplification, its global topology, and its long-term behavior
\cite{Kiuchi2024arXiv240510081K, RadiceHawke2024LRCA...10....1R}.
A growing number of general-relativistic magnetohydrodynamic (GRMHD) simulations have investigated various aspects of the impact of magnetic fields in these events \cite{Ruiz_etal2016ApJ...824L...6R, Ciolfi_etal2019PhRvD.100b3005C, Palenzuela_etal2022PhRvD.106b3013P, AguileraMiret_etal2022ApJ...926L..31A, AguileraMiret_etal2023PhRvD.108j3001A,
Kiuchi_etal2014PhRvD..90d1502K, CombiSiegel2023PhRvL.131w1402C, Kiuchi_etal2015PhRvD..92l4034K, Kiuchi_etal2018PhRvD..97l4039K, Kiuchi_etal2023PhRvL.131a1401K, Kiuchi_etal2024NatAs...8..298K, Ruiz_etal2011PhRvD..83b4025R, Ruiz_etal2020PhRvD.101f4042R, Ruiz_etal2021PhRvD.104l4049R, Chabanov_etal2023ApJ...945L..14C, 
Ciolfi2020MNRAS495L..66C, 
Kawamura_etal2016PhRvD..94f4012K,
CiolfiKalinani2020ApJ...900L..35C}.
While these studies have reached consensus on certain outcomes, significant discrepancies and uncertainties persist across simulations.
One of the open questions is to what extent the commonly adopted, ad-hoc initial magnetic field configurations may affect the post-merger evolution.
Previous studies have shown that the orientation, strength, or topology of pre-merger magnetic fields can impact jet launching \cite{Ruiz_etal2018PhRvD..98l3017R, Ruiz_etal2020PhRvD.101f4042R}, imprint features in the GW signal \cite{Giacomazzo_etal2009MNRAS.399L.164G, Ruiz_etal2021PhRvD.104l4049R, Bamber_etal2024PhRvD.110b4046B, Bamber_etal2025PhRvD.111d4038B}, and affect magnetic field amplification \cite{Chabanov_etal2023ApJ...945L..14C, Kawamura_etal2016PhRvD..94f4012K}.
On the other hand, some works that use subgrid turbulence modeling suggest that the pre-merger field configuration is largely erased after the merger \cite{AguileraMiret_etal2022ApJ...926L..31A}, and a similar universal amplification pattern would occur across different EOSs or mass ratios \cite{AguileraMiret_etal2025arXiv250410604A}.

In this work, we present a suite of GRMHD simulations performed with {\tt GR-Athena++} \cite{Stone_etal2020ApJS..249....4S, Daszuta_etal2021ApJS..257...25D, Cook_etal2025ApJS..277....3C} to study the evolution of magnetic fields in BNS mergers.
We considered different binary setups varying the initial magnetic field configuration, including purely poloidal fields with different orientations, purely toroidal fields, mixed poloidal/toroidal fields, and a setup where we imposed reflection (bitant) symmetry across the equatorial plane.
In addition, we explored two different mass ratios and two different equations of state (EOS).
We focus on the post-merger time and spatial evolution of various quantities associated with magnetic and thermodynamic properties, differentiating the evolution within the post-merger remnant and the surrounding disk.
In a companion paper, Cook {\it et al.} (in prep.) \cite{Cook_etal2025_inprep}, from now on Paper II, we will investigate mass ejection, gravitational wave emission, and global magnetic field amplification.

The remainder of the paper is organized as follows.
In Sec. \ref{sec:methods}, we describe the numerical setup, initial conditions, and binary configurations for our set of simulations.
In Sec. \ref{sec:results}, we present our main results.
We first describe the qualitative dynamics of the binary mergers by visualizing the structure of the magnetic field and other quantities.
Then, we perform a more quantitative analysis of the time evolution and spatial distribution of several properties associated with the thermodynamic and magnetic properties of the system.
In Sec. \ref{sec:discussion}, we compare our findings with previous works and discuss the broader implications of our results.
Finally, in Sec. \ref{sec:conclusions}, we summarize our key conclusions and outline future research directions.

\section{Methods}
\label{sec:methods}

\begin{table*}
    \centering
    \begin{tabular}{cccccccccc}
        \hline
        {\bf Simulation} & {\bf Res.}  & $m_1~[M_\odot]$  & $m_2~[M_\odot]$  & {\bf EOS} & $B$ {\bf config.} &  $\mathcal{B}_{\rm max}^{t=0}~[10^{15}{\rm G}]$ & $t_{\rm end}-t_{\rm mer}~[{\rm ms}]$ & $M_{\rm disk}~[M_\odot]$ & $M_{\rm ej}~[M_\odot]$ \\
        \hline 
        \POLUU      &   SR    & $1.35$  & $1.35$ & SFHo & Pol. Aligned & $5$ & $28.1$ & $8.5 \times 10^{-2}$ & $2.68 \times 10^{-2}$ \\
        \POLUD   & SR  & $1.35$ & $1.35$ & SFHo & Pol. Antialigned & $5$ & $31.7$ & $9.9 \times 10^{-2}$ &   $2.26 \times 10^{-2}$\\
        \POLLR & SR  & $1.35$ & $1.35$ & SFHo & Tilted dipole ($90^\circ$) & $5$ & $21.3$ & $1.01 \times 10^{-1}$ &   $1.93 \times 10^{-2}$\\
        \POLASYM  & SR  & $1.35$ & $1.35$ & SFHo & Pol. displaced & $5$ & $31.1$ & $9.3 \times 10^{-2}$  &   $1.88 \times 10^{-2}$\\
        \TORPP &  SR   & $1.35$ & $1.35$ & SFHo & Tor. Aligned & $5$ & $25.2$ & $9.6 \times 10^{-2}$  &   $2.22 \times 10^{-2}$\\
        \TORPM &  SR  & $1.35$ & $1.35$ & SFHo & Tor. Antialigned & $5$ & $28.1$ & $7.9 \times 10^{-2}$  &   $3.26 \times 10^{-2}$\\
        \MIX  & SR  & $1.35$ & $1.35$ & SFHo & Mixed Pol./Tor. & $5$ & $29.1$ & $1.03 \times 10^{-1}$  &   $2.71 \times 10^{-2}$\\
        \BITANT  & SR  & $1.35$ & $1.35$ & SFHo & Pol. (Bitant) & $5$ & $17.6$ & $9.5 \times 10^{-2}$  &   $2.50 \times 10^{-2}$\\
        \POLQ  & SR   & $1.47$ & $1.22$ & SFHo & Pol. Aligned & $5$ & $33.4$ & $1.4 \times 10^{-1}$  &   $0.963 \times 10^{-2}$\\
        \STIFF    &    SR      & $1.35$ & $1.35$ & DD2 & Pol. Aligned & $5$ & $23.3$ & $1.26 \times 10^{-1}$  & $2.04 \times 10^{-2}$ \\
        \LOWB  &  SR   & $1.35$ & $1.35$ & SFHo & Pol. Aligned & $5\times 10^{-7}$ & $39.1$ & $1.3 \times 10^{-1}$  &   $2.33 \times 10^{-2}$\\
        \NOB   &  SR  & $1.35$ & $1.35$ & SFHo & Hydro & $0$ & $24.2$ & $9.9 \times 10^{-2}$  &   $2.56 \times 10^{-2}$\\
        \HRPOLUU   &  HR  & $1.35$ & $1.35$ & SFHo & Pol. Aligned & $5$ & $13.2$ & $8.6 \times 10^{-2}$   &   $2.40 \times 10^{-2}$\\
        \HRMIX   &  HR  & $1.35$ & $1.35$ & SFHo & Mixed Pol./Tor. & $5$ & $15.1$ & $6.4 \times 10^{-2}$  &   $2.27 \times 10^{-2}$\\
        \hline 
    \end{tabular}
    \caption{List of simulations considered in this study. From left to right we list the assigned name of the simulation, the NS component gravitational masses ($m_1$, $m_2$), the EOS, the initial configuration of the magnetic field, the initial maximum value of the magnetic field ($\mathcal{B}_{\rm max}^{t=0}$), the duration of the simulation after merger ($t_{\rm end}-t_{\rm mer}$), and the disk ($M_{\rm disk}$) and ejecta ($M_{\rm ej}$) masses at the end of each simulation. For the poloidal configurations, by `aligned' (`antialigned') we refer to parallel (antiparallel) magnetic moments in both stars, whereas for the toroidal configurations, this term refers to the direction (counterclockwise or clockwise) of the field in both stars. The tilted dipole corresponds to the same topology as in the poloidal runs, but with the magnetic moment rotated $90^\circ$ and thus lying on the plane}.
    \label{tab:sims}
\end{table*}

  The binary system's initial configuration consists of two neutron stars orbiting in a quasi-circular orbit.
  All but one of our simulations consist of equal-mass NSs with gravitational masses $m_1=m_2=1.35 ~M_\odot$ at an initial separation of $40$ km; the remaining simulation (\POLQ) evolves an unequal mass binary with NS masses $m_1=1.47~M_\odot$ and $m_2=1.22~M_\odot$ (a mass ratio of $q=1.2$) at an initial separation of $45$ km.
  The initial data was obtained using the {\tt Lorene} pseudo-spectral code \cite{Gourgoulhon_etal2001PhRvD..63f4029G}, and all but one of our simulations assume the soft SFHo \cite{Steiner_etal2013ApJ...774...17S} EOS to describe the nuclear properties of the high-density matter; the other simulation (\STIFF) assumes the DD2 EOS \cite{Typel_etal2010PhRvC..81a5803T, HempelSchaffnerBielich2010NuPhA.837..210H}.
  
  On top of the hydrodynamic initial data solution, we superimpose a magnetic field confined within each of the two stars.
  We consider purely poloidal dipole-like fields, with different orientations of the magnetic moments, purely toroidal fields with varying polarities, and a mixed poloidal/toroidal configuration.
  In the latter case, we follow the results in Ref. \cite{Cook_etal2025arXiv250608037C} and assume that $85\%$ of the magnetic energy is in the poloidal field and $15\%$ of it is in the toroidal field.
  For the poloidal cases, we define a purely toroidal vector potential of the form
\begin{equation}
    \textbf{A} = A_{\rm b} \max(p-p_{\rm cut},0)^{n_{\rm s}} (-y, x, 0),
\end{equation}
with $n_{\rm s}=3$, $p_{\rm cut}=0.1p_{\rm max}$, where $p_{\rm max}$ is the maximum gas pressure of the star, $x$ and $y$ are measured from the center of each star.
Then, if needed, we apply a rotation of the vector potential to account for a tilted dipolar field.
For the toroidal magnetic field, we choose a vector potential of the form \cite{Hayashi_etal2023PhRvD.107l3001H, Bamber_etal2024PhRvD.110b4046B}
\begin{multline}
\textbf{A} = A_{\rm b} \max(p - p_{\mathrm{cut}},0)^{n_s} \\ \times \left[ x(z^2-R^2),y(z^2-R^2),-z(x^2+y^2-R^2) \right],
\label{eq:toroidalA}    
\end{multline}
where $R$ is the radius of the stars along the axis of the binary.
  The normalization constant $A_{\rm b}$ is chosen so that the maximum (conserved) magnetic field, $\mathcal{B}^i:=\sqrt{\gamma}B^i$, is initially $\mathcal{B}_{\rm max}(t=0)= 5\times 10^{15}~{\rm G}$ and it is located at the center of the stars; here, $B^i$ are the components of the magnetic field measured by an Eulerian observer and $\gamma:=\det({\gamma_{ij}})$ is the determinant of the spatial metric $\gamma_{ij}$.
  For comparison, we also evolve a binary without magnetic fields (\NOB) and one with a very low magnetic field of $\mathcal{B}=5\times 10^8~{\rm G}$ (\LOWB).
  In our magnetized cases, the initial magnetic energy, calculated as
\begin{equation}
  E_B = \frac{1}{2}\int_{\Sigma_t} \sqrt{\gamma} W b^2 d^3{\bf x},
\end{equation}
where $b^2= b^\mu b_\mu$, $b^\mu$ is the magnetic field 4-vector\footnote{The definition of $b^\mu$ includes a factor $(4\pi)^{-1/2}$.} in the fluid frame and $W$ is the Lorentz factor of the fluid in the Eulerian frame (for the details on these definitions, see Ref. \cite{Cook_etal2025ApJS..277....3C}), is $E_B \approx 5 \times 10^{47}~{\rm erg}$.

  We use the GRMHD code {\tt GR-Athena++} \cite{Stone_etal2020ApJS..249....4S, Daszuta_etal2021ApJS..257...25D, Cook_etal2025ApJS..277....3C, Daszuta_etal2024arXiv240609139D} which evolves magnetized fluids in dynamic spacetimes, employing a shock-capturing finite-volume scheme.
  A subset of our simulations uses the $5$th order WENOZ reconstruction method, while the rest of the runs use the piecewise parabolic method (PPM), and all of them use a local Lax--Friedrichs (LLF) Riemann solver.
  In Paper II, we discuss the effects that can be associated with the different reconstruction methods used.
  Magnetic fields are defined on cell faces and evolved with the constrained transport method \cite{GardinerStone2005JCoPh.205..509G, GardinerStone2008JCoPh.227.4123G} that maintains the divergence-free constraint down to machine precision.
  The spacetime is dynamically evolved using the Z4c formalism \cite{BernuzziHilditch2010PhRvD..81h4003B, Ruiz_etal2011PhRvD..83b4025R, Weyhausen_etal2012PhRvD..85b4038W, Hilditch_etal2013PhRvD..88h4057H} with the standard ``moving puncture gauge'' \cite{BrandtBrugmann1997PhRvL..78.3606B, Campanelli_etal2006PhRvL..96k1101C, Baker_etal2006PhRvL..96k1102B}; the metric quantities are vertex-centered and then interpolated to the cell-centers to be coupled with the fluid, and viceversa.
  The simulations are conducted on a cartesian grid with a length of ${\approx} 3020~{\rm km}$, a base resolution of $128^3$ points, and $7$ levels of refinement down to the inner region of size ${\approx} 74 \times 74 \times 37 ~ {\rm km}^3$, reaching a maximum resolution of $180~{\rm m}$ (from now on, standard resolution or SR).
 We also conducted two high-resolution (HR) calibration runs for the poloidal and mixed magnetic field configurations, reaching a maximum resolution of ${\sim} 123~{\rm m}$.
  
  Table \ref{tab:sims} lists the main setup quantities for all the simulations considered in this study, as well as outcomes such as the final ejecta (see Paper II) and disk masses.
  Note that the values of these latter quantities are affected by the duration of the simulation, especially for the shortest ones.

\section{Disk and remnant evolution}
\label{sec:results}

In this section, we present the results of our analysis of the set of BNS simulations.
We first present a qualitative description of the dynamics of the mergers and discuss the main differences among the various setups considered.
Then, we quantitatively analyze the evolution of several quantities of interest through the calculation of mass histograms, radial profiles, and the time evolution of averaged values within the remnant and the disk.
In what follows, we identify the ``remnant'' as the region which contains bound material ($-u_t < 1$) with densities $\rho > 10^{13}~{\rm g~cm}^{-3}$, and the ``disk'' as the region containing bound material with densities in the range $2 \times 10^8 < \rho < 10^{13}~{\rm g~cm}^{-3}$.

\subsection{Qualitative dynamics}

\begin{figure*}
  \centering
  \includegraphics[width=0.99\textwidth]{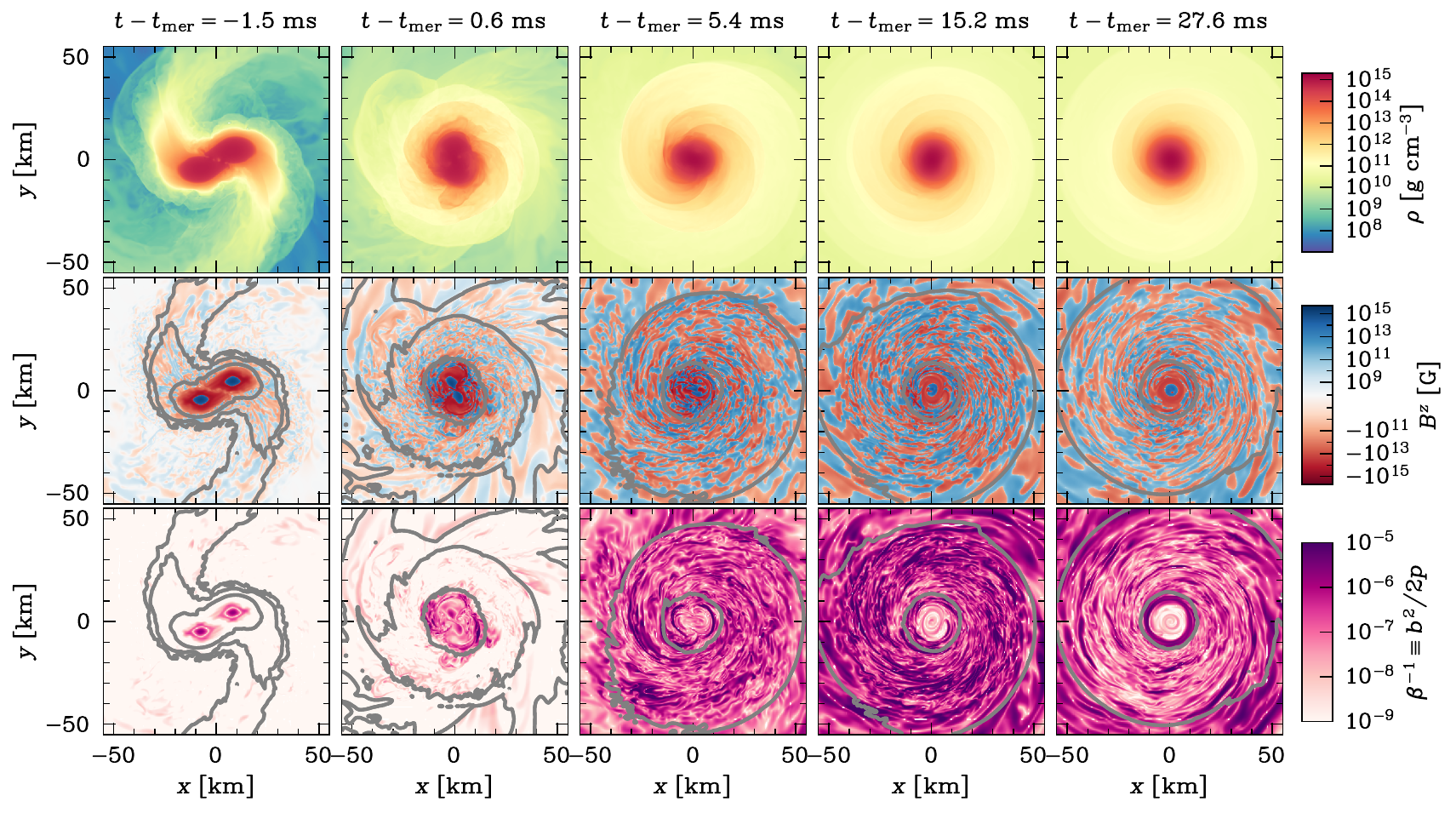}
  \caption{Slices of various quantities on the equatorial plane for the \POLUU\ simulation at different times relative to merger time.
 The panels display the evolution of (top to bottom) the rest-mass density ($\rho$), the vertical component of the magnetic field ($B^z$), and the inverse of the plasma $\beta$ parameter.
 In the bottom two panels, isodensity contours are shown in gray for $\rho=5\times 10^9, 5\times 10^{10},~{\rm and}~ 10^{13}~{\rm g~cm}^{-3}$.
 Each column corresponds to a different time: ($t-t_{\rm mer}=-1.5,0.5,5, 15, 27.6~{\rm ms}$).}
  \label{fig:fiducial_xy}
\end{figure*}

\begin{figure*}
  \centering
  \includegraphics[width=0.997\textwidth]{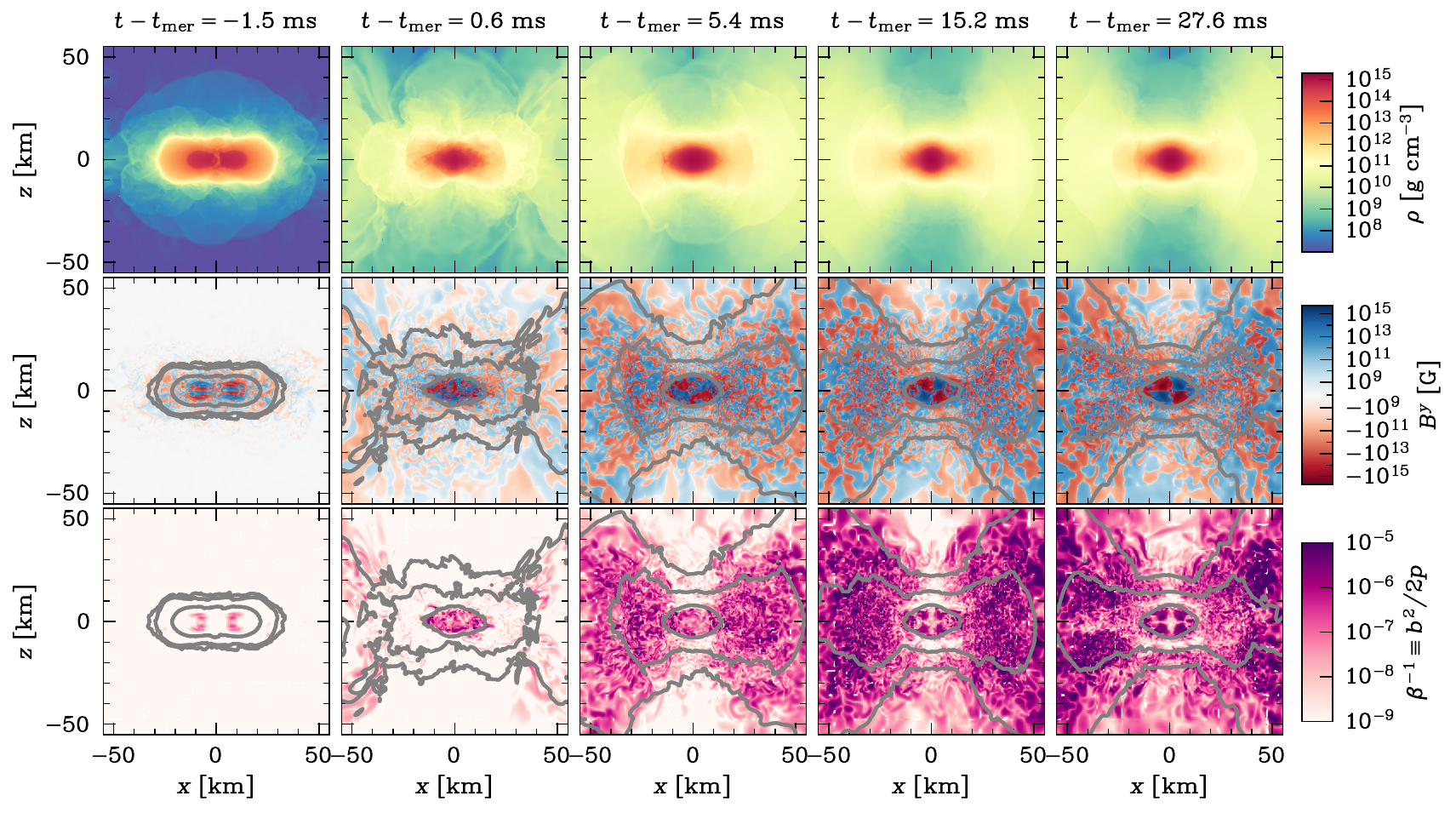}
  \caption{Slices of various quantities on the $xz$ plane for the \POLUU\ simulation at different times relative to merger time.
 The panels display the evolution of (top to bottom) the rest-mass density ($\rho$), the magnetic field component perpendicular to the plane ($B^y$), and the inverse of the plasma $\beta$ parameter.
 In the bottom two panels, isodensity contours are shown in gray for $\rho=5\times 10^9, 5\times 10^{10},~{\rm and}~ 10^{13}~{\rm g~cm}^{-3}$.
 Each column corresponds to a different time: ($t-t_{\rm mer}=-1.5,0.5,5, 15, 27.6~{\rm ms}$).}
  \label{fig:fiducial_xz}
\end{figure*}

Binaries with an initial separation of $40$ ($45$) km perform $\approx 3$ ($6$) orbits before merging.
We define the merger time as the retarded time for the maximum of the GW $(2,2)$ strain amplitude: $t_{\rm mer}:=t_{\rm max,GW} - R_{\rm GW}/c$, where $R_{\rm GW}\simeq 442~{\rm km}$ is the radius at which GWs are recorded and $c$ is the speed of light.
Equal-mass binaries with the SFHo EOS merge at a time of $t_{\rm mer}\simeq 10.3~{\rm ms}$ when using PPM and $t_{\rm mer}\simeq 11.1~{\rm ms}$ when using WENOZ, whereas the \STIFF\ run, with the DD2 EOS, merges at $t_{\rm mer}\simeq 9~{\rm ms}$.
On the other hand, simulation \POLQ\ merges at $t_{\rm mer}\simeq 15.9~{\rm ms}$, because of the larger initial separation.
Depending on the simulation, we continued the evolution for ${\sim} 15{-}35~{\rm ms}$ after the merger.
All mergers give birth to massive remnants, which survive against gravitational collapse for the entire duration of the runs.
We remark that for the masses considered, the binaries with the SFHo EOS are expected to produce a black hole at some time after the merger.
Other simulations with a similar setup but performed with a different code have produced a gravitational collapse in a range $\sim 5-25~{\rm ms}$ after the merger (e.g., Refs. \cite{Radice_etal2018ApJ...869..130R, CombiSiegel2023ApJ...944...28C}).
Nevertheless, the collapse time depends on the resolution, as well as on viscous and other dissipative effects, and thus it is not a reliable prediction from simulations \cite{Zappa_etal2023MNRAS.520.1481Z}.

Figure \ref{fig:fiducial_xy} shows equatorial slices ($z=0$) at different times for simulation \POLUU, illustrating the time-evolution of representative plasma quantities.
From top to bottom, the panels show the rest-mass density ($\rho$), the vertical magnetic field ($B^z$), and the magnetic pressure to gas pressure ratio, $ \beta^{-1} \equiv p_B/p_{\rm gas}$, where $p_B=b^2/2$\footnote{In some parts of the paper, we refer to this quantity as the `magnetization' because its behavior is very alike that of the $\sigma:=b^2/\rho$, but it must not be confused with this latter quantity.}.
From left to right, each column corresponds to an increasing time relative to the merger time: $t - t_{\rm mer} = -1.5$, $0.5$, $5$, $15$, and $27.6$ ms (last snapshot for this particular simulation).
In turn, Figure \ref{fig:fiducial_xz} shows equivalent 2D slices but for the $y=0$ plane; in this case, the middle row shows the toroidal component of the magnetic field ($B^y$).

When the two NSs merge, an oblate non-axisymmetric massive remnant rapidly forms.
At the same time, gas is expelled by dynamical forces; some of it unbinds and escapes, forming the so-called dynamical ejecta, while the rest remains bound and starts forming a disk around the remnant.
The expelled gas advects the magnetic field initially confined within the stars and spreads it out of the remnant; this occurs predominantly close to the equatorial plane.
Figure \ref{fig:fiducial_xy} shows this effect: initially, the vertical field is predominantly coherent and poloidal, and it is confined within the stars.

The first magnetic field structures that emerge from the remnant are those attached to the material in the spiral arms formed in the late inspiral and early merger (see left panel of Fig. \ref{fig:fiducial_xy}).
The initially coherent vertical field rapidly spreads out and becomes turbulent.
On the contrary, the toroidal field (middle row in Fig. \ref{fig:fiducial_xz}) starts from a turbulent state before the merger and slowly becomes more ordered later, especially within the remnant.

In the post-merger phase, the toroidal field develops predominantly different polarities above and below the plane; namely, the toroidal field below the plane is approximately a reflection of the field above the plane.
However, this effect is not perfectly symmetric across the equatorial plane.
In the core of the remnant, a field with negative polarity (counter-clockwise) that dominates above the equatorial plane extends to the regions below the plane, and the opposite happens in the outer regions of the remnant.
This can be seen in the middle row of Fig. \ref{fig:fiducial_xz}.

The evolution of $\beta^{-1}$ reflects both the amplification and redistribution of the magnetic field.
The magnetic field is amplified within the remnant shortly after the merger due to the KH instability (see Paper II).
Part of this amplified field is then expelled from the remnant and populates the disk, increasing its magnetic-to-gas pressure ratio there.
In the late stages of the simulation, the largest magnetization is present in the outer regions of the disk at moderate angles from the equatorial plane.
The bottom row in Fig. \ref{fig:fiducial_xz} also shows that the polar funnel is still dominated by gas pressure by the end of the simulation, indicating that this region is still heavily polluted by baryons.
This is consistent with the findings in Ref. \cite{Ciolfi_etal2019PhRvD.100b3005C}, who showed that this pollution prevents (or delays) the formation of a relativistic jet.
Nevertheless, our treatment neglects neutrino transport, which has been demonstrated to contribute, among other effects, to decreasing the baryon pollution in the funnel and then help a jet to form \cite{Moesta_etal2020ApJ...901L..37M, CombiSiegel2023PhRvL.131w1402C, Musolino_etal2025ApJ...984L..61M}.

\begin{figure*}
  \centering
  \includegraphics[width=0.9999\linewidth]{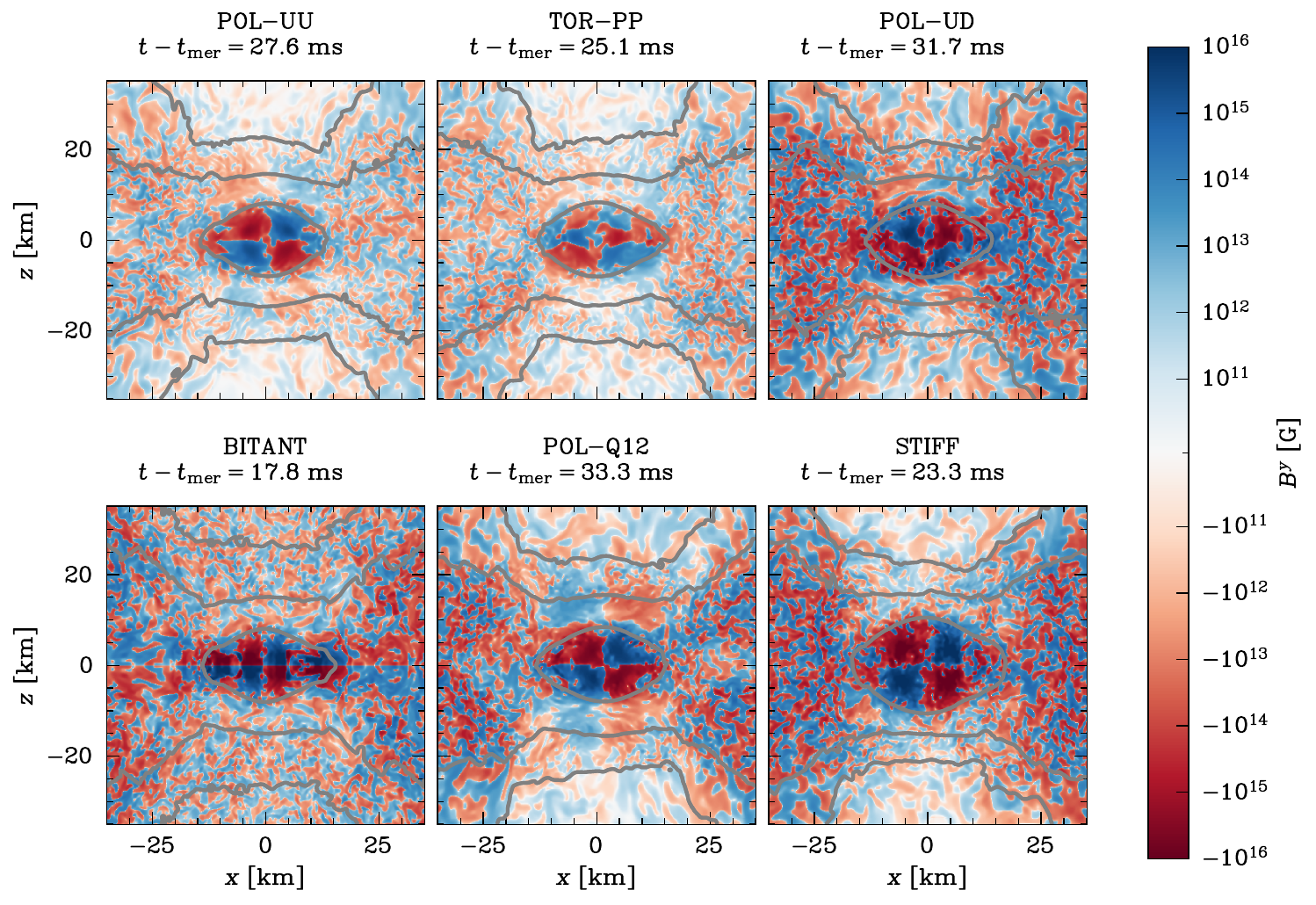}
  \caption{Vertical slices ($y=0$) for the last snapshot of a subset of the simulations displaying the component of the magnetic field perpendicular to the plane, $B^y$.}
  \label{fig:fiducial_xz_all}
\end{figure*}

To compare the effect of the initial binary configuration and magnetic field on the late magnetic structure of the remnant and the disk, we show in Fig. \ref{fig:fiducial_xz_all} a closer look at the toroidal magnetic field component in the $y=0$ plane for the last snapshot of a subset of our simulations.
We have indicated above each panel the time of the last snapshot for each one.
Simulations with an initially vertical and aligned dipolar field give rise to a toroidal field that follows the natural twisting caused by the rotation of the remnant, namely, this produces a counter-clockwise (clockwise) directed field in the regions above (below) the equatorial plane.
This is the case for \POLUU, \BITANT, \POLQ, and \STIFF, which share the same initial field configuration.
On the other hand, simulations with a different initial configuration, such as initially antialigned (\POLUD) dipole-like fields or toroidal fields (\TORPP), develop a toroidal component with the opposite direction above and below the plane compared with the aforementioned runs.
The \POLUD\ configuration is of particular interest, as antialigned fields do not prescribe a preferred direction for the post-merger toroidal component.
Accordingly, the topology we see in our simulation may not constitute a generic outcome of this setup, but rather reflect a stochastic realization influenced by subtle asymmetries in the initial conditions.

\begin{figure}
  \centering
  \includegraphics[width=0.997\linewidth]{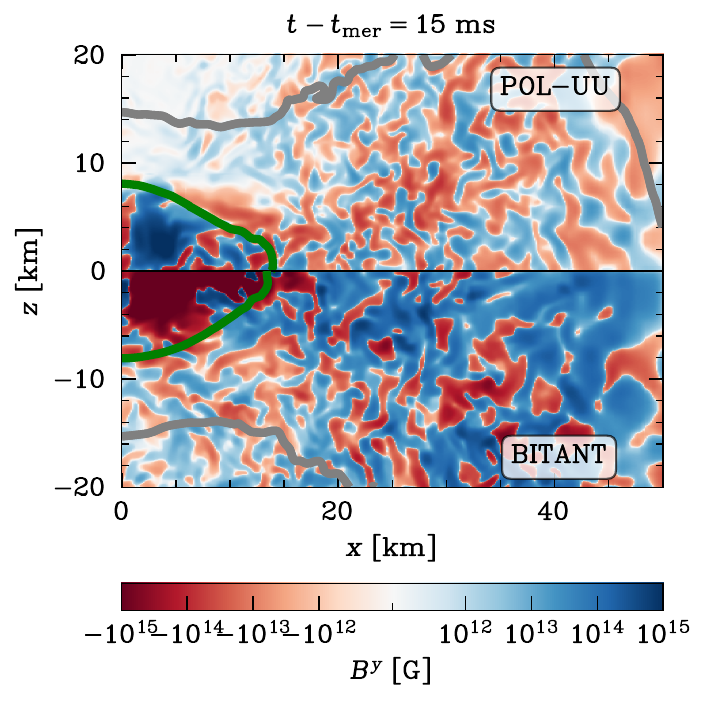}
  \caption{Vertical slices ($y=0$) at $t=15~{\rm ms}$ for \POLUU\ and \BITANT, displaying the component of the magnetic field perpendicular to the plane, $B^y$. The \BITANT\ simulation shows a lower development of turbulence near the equatorial plane due to the reflective boundary conditions.}
  \label{fig:fiducial_xz_zoom}
\end{figure}

Figure \ref{fig:fiducial_xz_zoom} shows a magnified comparison between the toroidal magnetic field in the \POLUU\ and \BITANT\ runs.
The simulation with bitant symmetry exhibits a more coherent structure near the equatorial plane, which is likely a consequence of the imposed boundary condition that forces the field to be perfectly symmetric.
This suggests that imposing reflective symmetry across the $z=0$ plane decreases the degree of field turbulence that develops close to the plane.

Simulation \POLQ\ ran for longer than the other runs and starts to show signs of ordering of the toroidal field in the funnel.
Finally, simulation \STIFF\ shows a more rapid ordering of the field within the remnant even at earlier times.
This is likely due to two effects: i) as a result of to the stiffness of the EOS, the remnant has a larger volume and thus more space for the formation of coherent structures, and ii) the collision is less violent than with the soft SFHo, and the initial coherent field survives the merger.

\begin{figure}
  \centering
  \includegraphics[width=0.9975\linewidth]{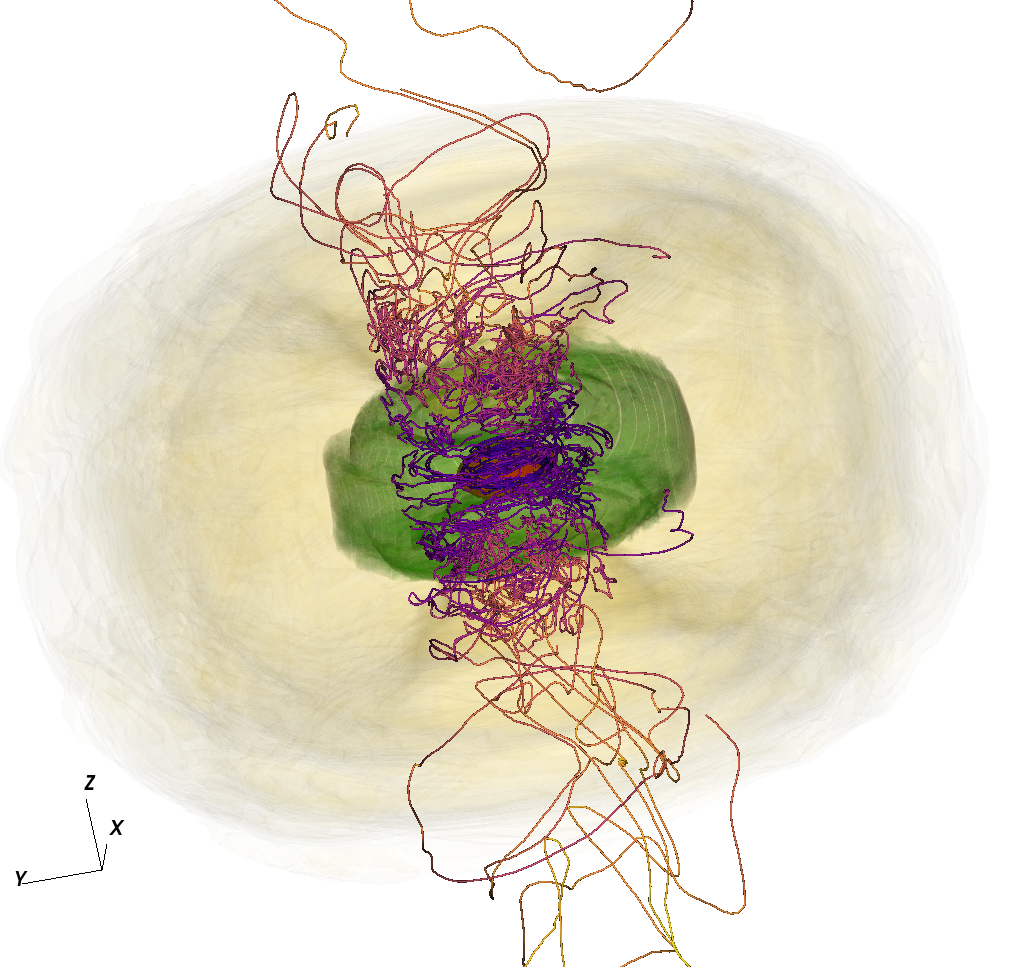}
  \caption{
  Three-dimensional rendering of rest-mass density and magnetic field structure for the remnant core and surrounding material of simulation \POLUD.
  The magnetic field lines are colored by strength, transitioning from yellow (weakest field) to purple (strongest field), emphasizing the twisted field structure in the post-merger remnant.
  The contour volumes show regions of different density, broadly corresponding to the remnant (red, $\rho > 10^{13}{\rm g~cm}^{-3}$), inner disk ($ 10^{12}{\rm g~cm}^{-3} < \rho < 10^{13}{\rm g~cm}^{-3}$), and outer disk ($ 2 \times 10^8{\rm g~cm}^{-3} < \rho < 10^{12}{\rm g~cm}^{-3}$).}
  \label{fig:3D_rendering}
\end{figure}

The global structure of the post-merger remnant, disk, and its associated magnetic field can be visualized in Fig.~\ref{fig:3D_rendering}, where we show a 3D rendering of the rest-mass density for a late-time simulation snapshot of simulation \POLUD, superposed with the magnetic field lines close to the vertical axes.
The red isodensity surfaces correspond to regions where $\rho \geq 10^{13}~{\rm g~cm}^{-3}$, which we defined as the post-merger remnant \cite{Hanauske_etal2017PhRvD..96d3004H}.
Beyond the core, we show in green isodensity surfaces with $ 10^{12}{\rm g~cm}^{-3} < \rho < 10^{13}{\rm g~cm}^{-3}$) and in yellow regions where $ 2 \times 10^8{\rm g~cm}^{-3} < \rho < 10^{12}{\rm g~cm}^{-3}$; these two regions correspond to the inner and outer disk.
The magnetic field structure is visualized using field lines colored by strength, transitioning from yellow (weakest field), through orange, red, and finally purple (strongest field).
At this stage of the simulations, the field lines begin to exhibit signs of the characteristic winding caused by the differential rotation of the remnant.

\subsection{Spatial and time evolution}
\label{sec:spatialandtime}

\subsubsection{Magnetothermal evolution}
\label{sec:thermo}

\begin{figure*}
  \centering
  \includegraphics[width=0.994\linewidth]{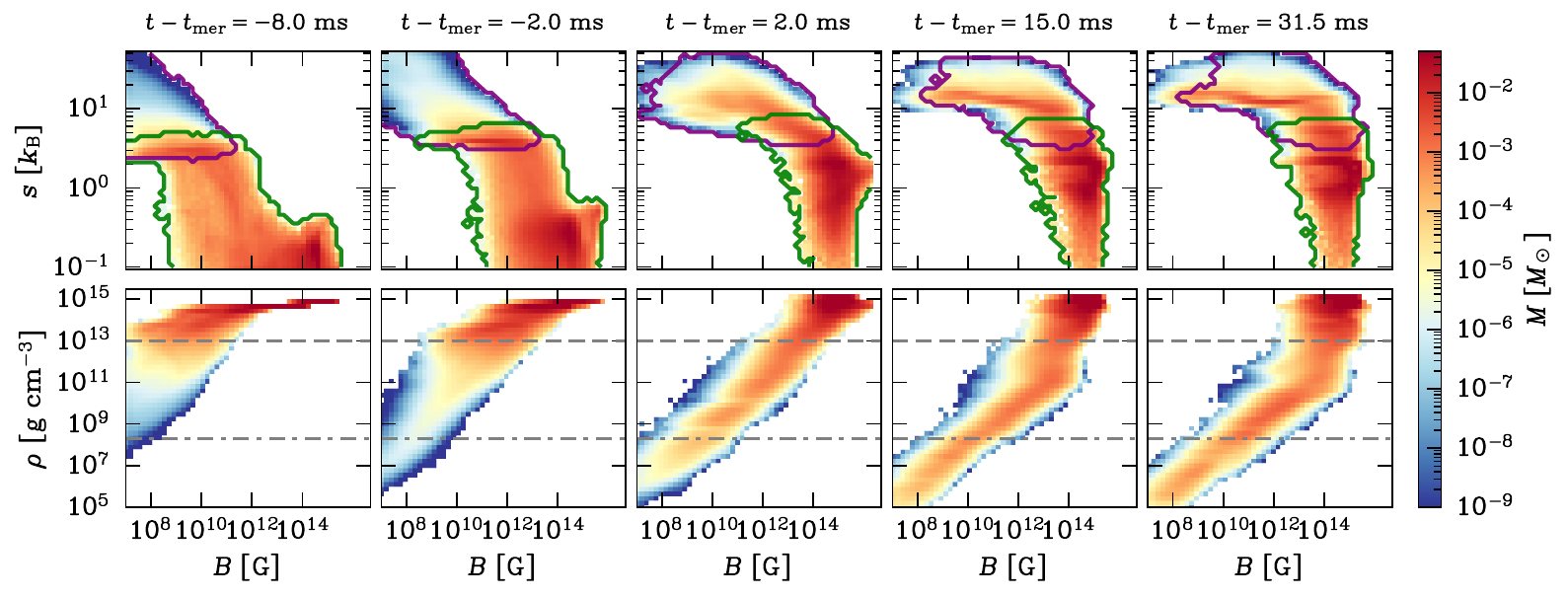}
  \caption{Two-dimensional mass-weighted histograms depicting the magnetic and thermodynamic evolution of the material in simulation \POLUD\ as a function of time. The top panel shows the mass evolution in the plane ($B$, $s$), and the bottom panel shows ($B$, $\rho$). From left to right, we show different times relative to the merger time.
  In the top panel, the mass within the remnant (disk) is enclosed within a purple (green) curve.
  In the bottom panel, the outer boundaries of the remnant and disk are marked by horizontal dashed and dot-dashed lines, respectively.}
  \label{fig:histograms}
\end{figure*}

\begin{figure*}
  \centering
  \includegraphics[width=0.994\linewidth]{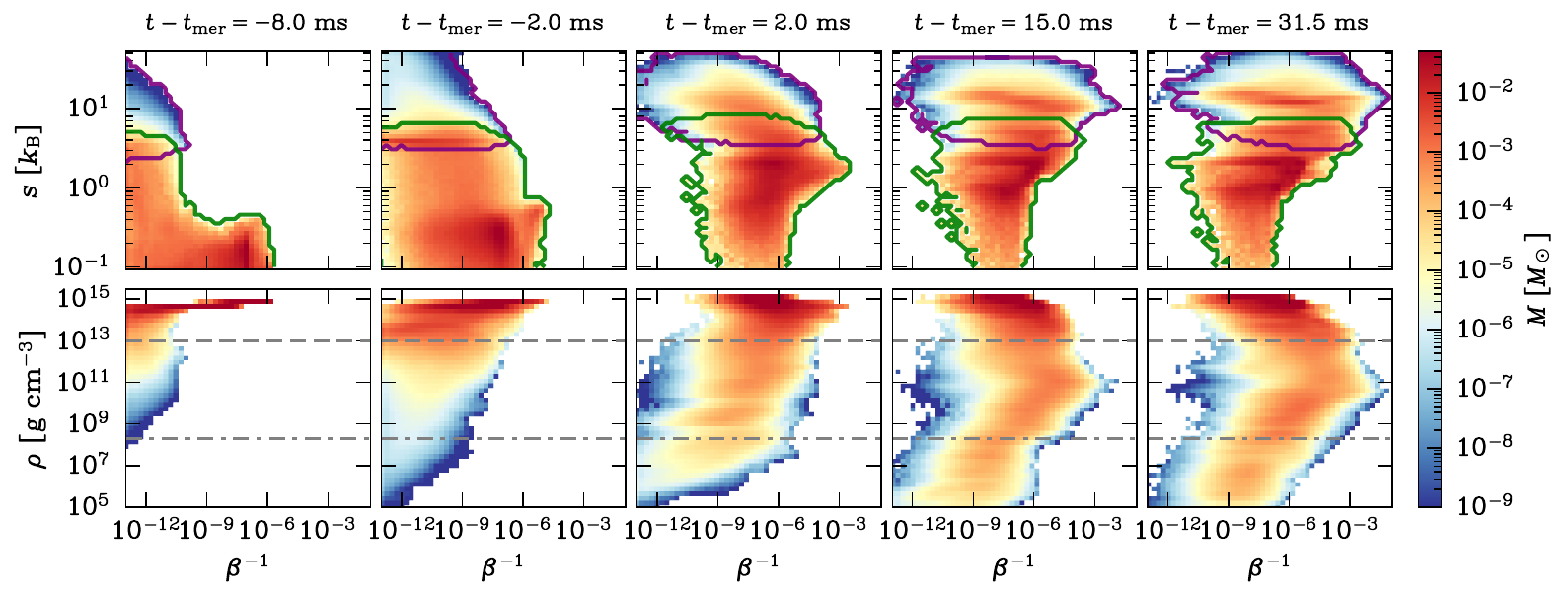}
  \caption{Two-dimensional mass-weighted histograms describing the magnetic and thermodynamic evolution of the material in simulation \POLUD\ as a function of time. The top panel shows the mass evolution in the plane ($\beta^{-1}$, $s$) and the bottom panel shows ($\beta^{-1}$, $\rho$). From left to right, we show different times relative to the merger time.
  In the top panel, the mass within the remnant (disk) is enclosed within a purple (green) curve.
  In the bottom panel, the outer boundaries of the remnant and disk are marked by horizontal dashed and dot-dashed lines, respectively.}
  \label{fig:histograms_invbeta}
\end{figure*}

  We investigate the magnetic and thermodynamic evolution of the post-merger remnant and disk by calculating mass-weighted histograms of these properties.
  Figure \ref{fig:histograms} displays the mass evolution as a function of magnetic field intensity, entropy per baryon, and rest-mass density for simulation \POLUD.
  Each column corresponds to a different snapshot in time; from left to right: $t - t_{\rm mer} = -8$, $-2$, $2$, $15$, and $28.8$ ms.
  The top panel shows how the distribution of mass evolves in the ($B$, $s$) plane, whereas the bottom panel shows the evolution in the ($B$, $\rho$) plane.
  In the top panels, the mass corresponding to the remnant (disk) material is enclosed with a green (purple) line.
  The boundaries between these regions in the bottom panel are marked with horizontal dashed and dot-dashed lines of constant $\rho$, respectively.
  In Fig. \ref{fig:histograms_invbeta}, we show equivalent histograms but where the magnetic-to-gas pressure ratio $\beta^{-1}$ is used instead of $B$ to characterize the magnetic evolution of the material.

Before the merger (left two panels), most of the mass is confined to the interior of the NSs at high densities, and the matter has low entropy per baryon ($s/k_{\rm B} \ll 1$).
Very little material has already detached from the stars' atmosphere to form a proto-disk.
The magnetic field is large ($B\sim 10^{15}~{\rm G}$) only in the NSs' cores, and remains low throughout most of the NS volume; this is also reflected in the magnetic-to-pressure ratio.
Shortly after the merger, the magnetic field is rapidly amplified and the material within the remnant heats up, reaching temperatures of up to ${\sim} 30~{\rm MeV}$; most of the remnant material achieves a magnetic field intensity $\gtrsim 10^{14}~{\rm G}$ and $s/k_{\rm B}\gtrsim 1$.
As we described in the previous sections, a significant part of this hot magnetized material is expelled through the equatorial plane, producing outflows
and populating the disk.

Shortly after merger (${\sim} 2~{\rm ms}$), the magnetic field intensity and the rest-mass density show an approximately linear correlation with each other ($B \propto \rho$) across $\gtrsim 7$ orders of magnitude.
However, this trend begins to break down at later times: the magnetic field continues to increase in lower-density regions, and more material is pushed to larger $B$ values.
At ${\sim} 30~{\rm ms}$ after the merger, the above-mentioned linear correlation remains up to $\rho \sim 10^{11}~{\rm g~cm}^{-3}$, in the center of the disk. In contrast, at higher densities the magnetic field is rather constant with a value of $B\sim 10^{14}~{\rm G}$.
This effect is also reflected in the $(\rho, \beta^{-1})$ diagram (bottom panel of Fig. \ref{fig:histograms_invbeta}), where after the merger, $\beta^{-1}$ keeps increasing throughout the outer remnant and the disk.
Due to the outward decrease of pressure $dp/dr <0$, the constant magnetic field reflects in a magnetization that grows outwards down to densities of $\rho \sim 10^{11}~{\rm g~cm}^{-3}$, below which it starts to decrease again.

The entropy per baryon also keeps growing at a slow rate during the post-merger phase, though the largest increase occurs immediately after the merger.
This large increase is a consequence of the strong shocks produced in the collision interface, while the growth afterward may indicate further sustained shocks, for example, between the spiral arms in the disk, or magnetic reconnection.
By the end of the simulations, most of the mass in the disk has an entropy per baryon of $s/k_{\rm B} \gtrsim 10$, while the remnant entropy spans a range $s/k_{\rm B} \sim 0.1-5$, with most of the material concentrated at values $s/k_{\rm B} \sim 1-2$.

\begin{figure*}
  \centering
  \includegraphics[width=0.994\linewidth]{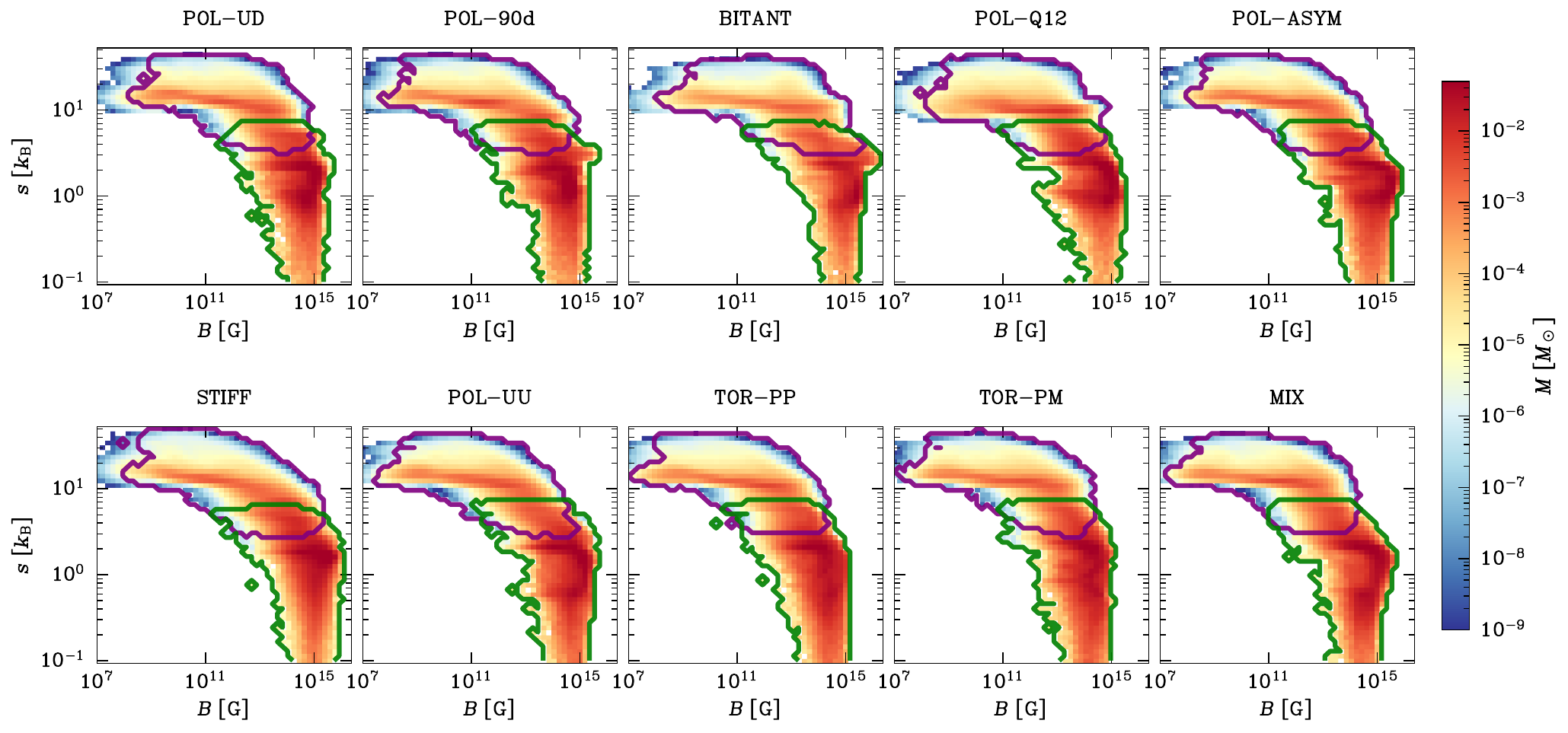}
    \caption{Two-dimensional mass-weighted histograms of the magnetic and thermodynamic state of the material at $t-t_{\rm mer}=15~{\rm ms}$ for each of the simulations. Each panel shows the mass evolution in the $(B, s)$ plane.
  The solid lines in the top row delimit the remnant and disk boundaries as defined in Fig. \ref{fig:histograms}.}
  \label{fig:histogram_all_B}
\end{figure*}

\begin{figure*}
\includegraphics[width=0.994\linewidth]{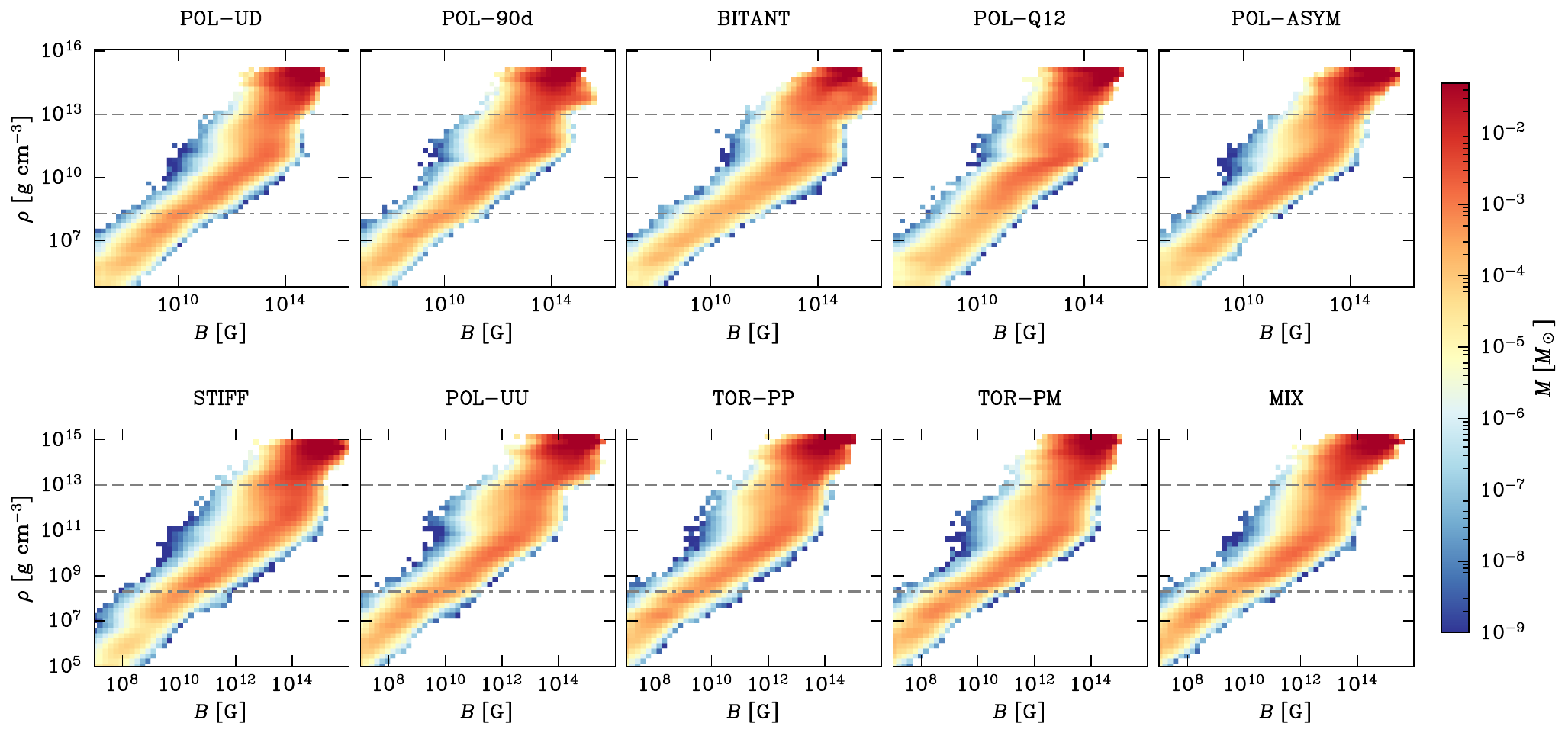}
  \caption{Two-dimensional mass histograms describing the magnetic state and thermodynamic evolution of the material at $t-t_{\rm mer}=15~{\rm ms}$ for each of the simulations. Each panel shows the mass evolution in the $(B, \rho)$ plane.
  The horizontal dashed lines delimit the remnant and disk boundaries as defined in Fig. \ref{fig:histograms}.}
  \label{fig:histograms_all_rho}
\end{figure*}

To compare the thermodynamic and magnetic field evolution of the different simulations, we show in Figs.~\ref{fig:histogram_all_B} and \ref{fig:histograms_all_rho} mass histograms for the last snapshot of each.
All simulations show similar behavior in their final snapshot, with small deviations, mainly resulting from the different times after the merger that each simulation reached.
In particular, in simulations interrupted at earlier times, the magnetic field is still growing in the inner disk and the outer remnant, and the relation between $\rho$ and $B$ is more complicated.
This difference should be alleviated at a later time.
It is not clear whether the magnetic field will keep increasing toward the saturation value of $B\sim 10^{14}~{\rm G}$ at densities lower than $10^{11}\ {\rm g}\ {\rm cm}^3$ over longer timescales.

\subsubsection{Time evolution}
\label{sec:time}

\begin{figure}
    \centering
    \includegraphics[width=0.985\linewidth]{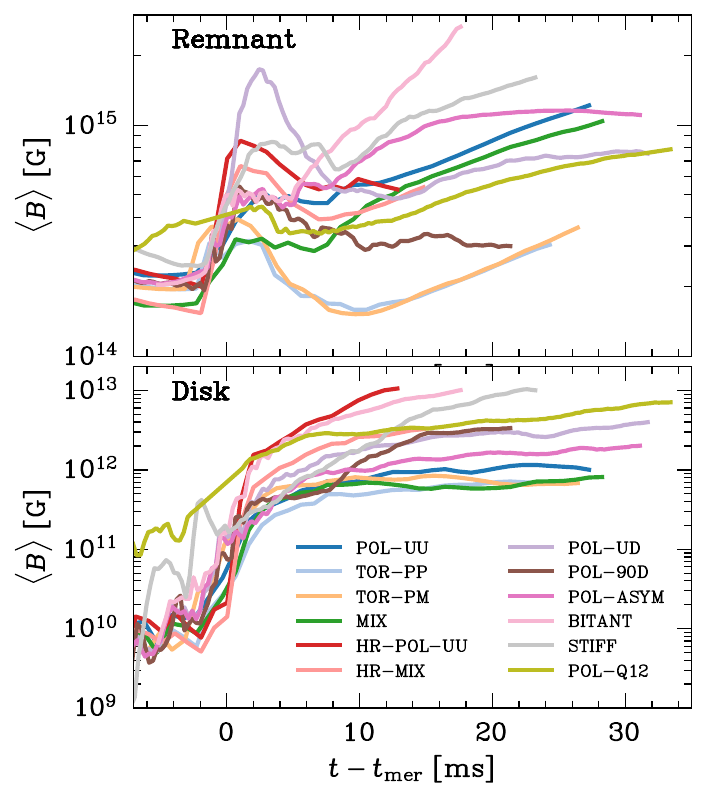}
    \caption{Average magnetic field in the remnant (top row) and the disk (bottom row) as a function of time for each of the simulations.}
    \label{fig:b_abs}
\end{figure}

\begin{figure}
    \centering
    \includegraphics[width=0.985\linewidth]{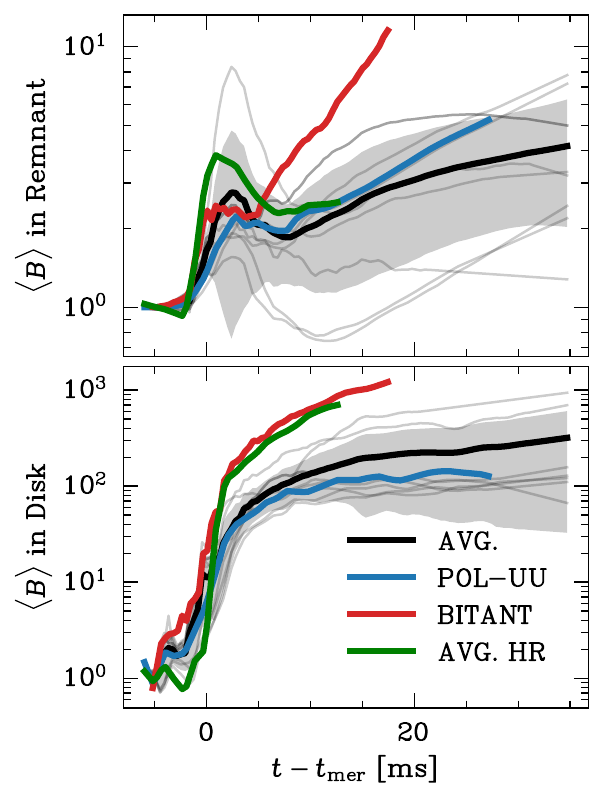}
    \caption{Average magnetic field as a function of time for runs with and without reflection symmetry. The black curve shows the average magnetic field for all the SR simulations without reflection symmetry, and the gray shaded area delimits one standard deviation to each side of the curve. The green curve shows the average magnetic field for the two HR simulations. In contrast, the blue and red curves show the average magnetic field for the \POLUU\ and the \BITANT\ run, respectively.}
    \label{fig:Bitant}
\end{figure}

\begin{figure}
    \centering
    \includegraphics[width=0.985\linewidth]{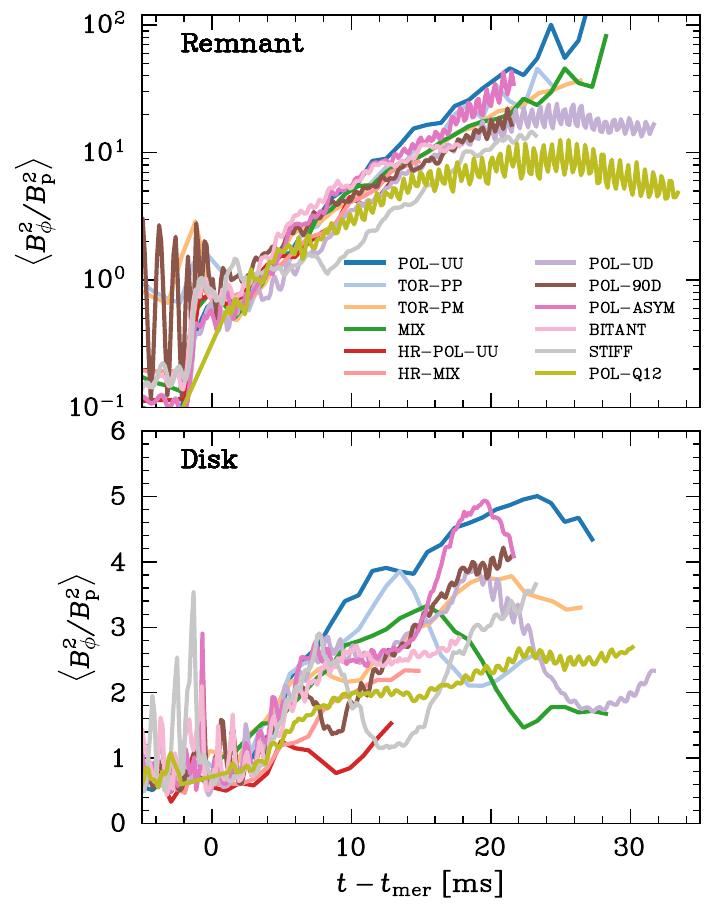}
    \caption{Ratio of the energy in the toroidal magnetic field ($B_\phi^2$) to the energy in the poloidal magnetic field ($B_{\bf p}^2$) integrated throughout the remnant (top row) and the disk (bottom row) as a function of time for each of the simulations. After the merger, the toroidal magnetic field dominates both in the remnant and the disk for all the simulations.}
    \label{fig:bphi2_bp2}
\end{figure}

\begin{figure}
    \centering
    \includegraphics[width=0.985\linewidth]{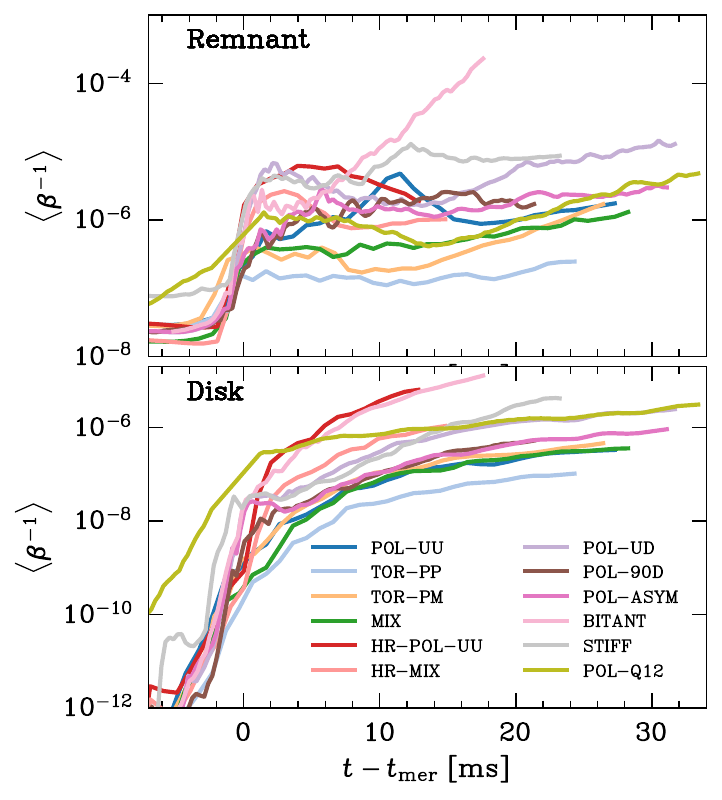}
    \caption{Average magnetic-to-gas pressure ratio ($\equiv \beta^{-1}$) in the remnant (top row) and the disk (bottom row) as a function of time for the various simulations.}
    \label{fig:avg_invbeta}
\end{figure}
\begin{figure}
    \centering
    \includegraphics[width=0.985\linewidth]{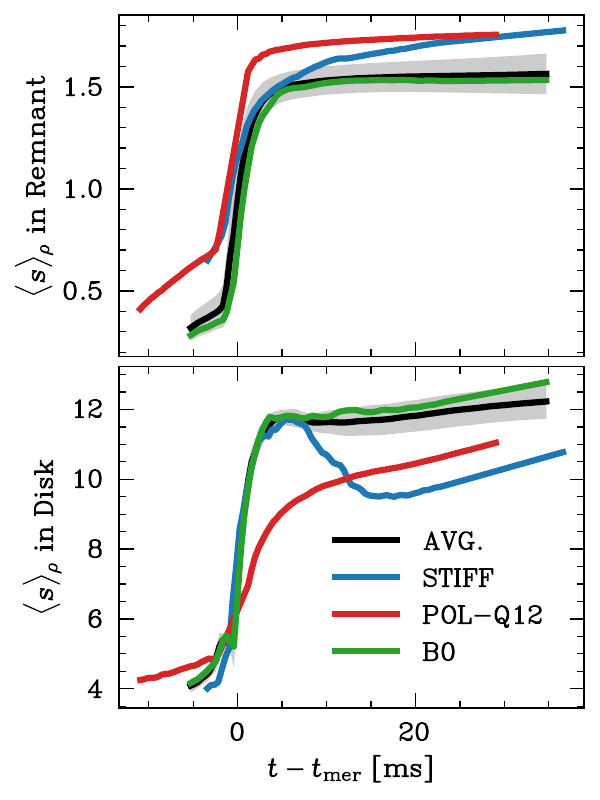}
    \caption{Average entropy per baryon in the remnant (top row) and the disk (bottom row) as a function of time for various simulations.
    The black curve is the averaged curve for all the simulations with equal mass NSs and the SFHo EOS, while the gray area encloses one standard deviation to each side of it.
    The blue and red curves correspond to the \STIFF\ and \POLQ\ simulations, respectively, which are the only ones showing significant differences with the other runs. The green curve shows the \NOB\ case, which shows no significant difference with the magnetized runs.}
    \label{fig:avg_s}
\end{figure}
\begin{figure}
    \includegraphics[width=0.985\linewidth]{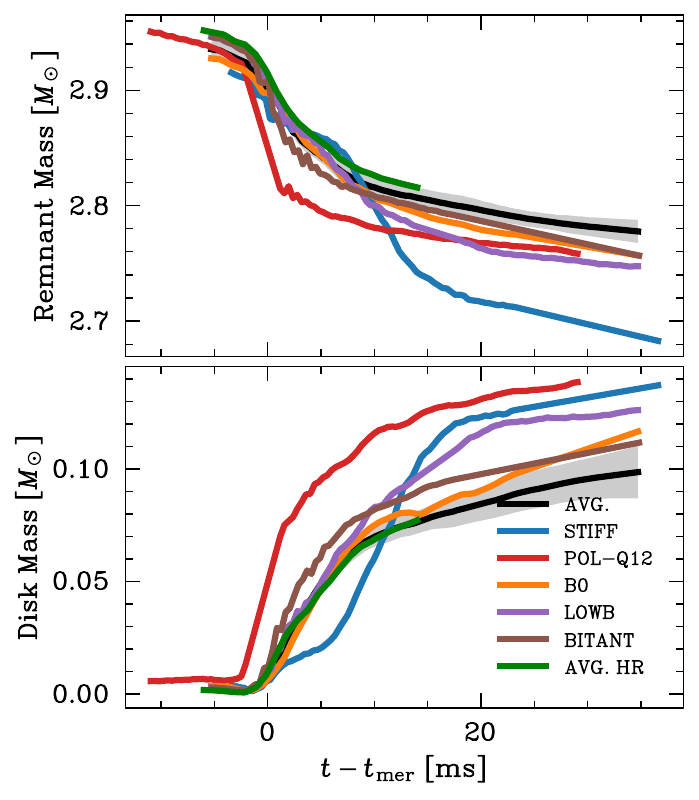}
    \caption{Baryonic mass in the remnant (top row) and the disk (bottom row) as a function of time for various simulations. The black curve shows the averaged mass for all the SR simulations with equal mass NSs, the SFHo EOS, and a large initial magnetic field, while the gray area encloses one standard deviation to each side of it. The green curve is the average mass for the two HR runs. The other curves correspond to simulations which a priori may show different disk and remnant masses: \STIFF, \POLQ, \NOB, \LOWB, and \BITANT.}
    \label{fig:mass}
\end{figure}

To have a more quantitative description of how the various properties of the system evolve in the remnant and the disk, we compute the time evolution of averaged quantities over these regions.
We compute the mean value of magnetic-field-related quantities $q$ (such as the magnetic-to-pressure ratio $\beta^{-1}$) as a volume averages,
\begin{equation}
    \langle q \rangle \equiv \frac{\int_V q \sqrt{\gamma}d^3{\bf x}}{\int_V \sqrt{\gamma}d^3{\bf x}},
\end{equation}
and the mean value of thermodynamic quantities (such as the entropy per baryon $s$) as a mass-weighted average,
\begin{equation}
    \langle q \rangle_\rho \equiv \frac{\int_V q Dd^3{\bf x}}{\int_V Dd^3{\bf x}}.
\end{equation}
In both cases, the integration is performed throughout the volume $V$ of the remnant ($\rho > 10^{13}~{\rm g~cm}^{-3}$) or the disk ($10^8~{\rm g~cm}^{-3} < \rho < 10^{13}~{\rm g~cm}^{-3}$).

Figure \ref{fig:b_abs} shows the time evolution of the average magnetic field value within the remnant (upper panel) and the disk (lower panel) for all our magnetized simulations.
In the remnant, there are large deviations among the different runs.
The exponential growth in the remnant due to the KH instability at the interface shear layer is a consistent outcome across the various setups, though there is a large variance in the maximum value of the magnetic field achieved.
After this phase, the magnetic field stabilizes for a short period or decreases, to then grow again at a lower rate due to turbulent amplification and finally due to winding.
In Paper II, we discuss in more detail the underlying mechanisms of the magnetic field amplification.
Here, we note that the \BITANT\ simulation shows a larger and much faster amplification than the other runs; this is reflected in Fig. \ref{fig:Bitant}, which shows how this run compares with the average magnetic field evolution of all the other runs, as well as with the \POLUU\ run, which has the same initial magnetic field but does not assume reflection symmetry across the equatorial plane.
It seems plausible that the imposed boundary condition--tending to keep the magnetic field $B$ predominantly poloidal on the equatorial plane--could facilitate a feedback process in which toroidal field generated by winding is, at least in part, converted back into poloidal components.
In this view, the winding mechanism draws a toroidal component from the poloidal field, while the boundary condition tends to restore a poloidal structure.
This is consistent with the amplification during the Kelvin--Helmholtz phase appearing similar in runs with and without bitant symmetry, with differences becoming more noticeable thereafter, when winding effects may start to play a larger role.

In the disk, the field starts to build up before the merger and continues growing after it when more magnetized material escapes from the remnant.
After the initial growing phase, the average magnetic field reaches a saturation, though there is a large variance among simulations ($\sim 5 \times 10^{11}~{\rm G} - 10^{13}~{\rm G}$).
The magnetic field evolution in the disk depends not only on the in situ amplification but also on the advection of the field from the remnant after the merger.
Though we do not resolve the fastest-growing MRI modes, unresolved MRI may play a minor role in the amplification of the field in the disk after the initial growing phase.
The \BITANT\ and \STIFF\ simulations have the most magnetized disks with magnetic field values ${\sim} 10^{13}~{\rm G}$, which are still slowly growing by the end of the simulations.

The evolution of the magnetic field topology can be addressed by comparing the toroidal and poloidal components of the magnetic field.
Figure \ref{fig:bphi2_bp2} shows the average of the toroidal-to-poloidal magnetic field energy ratio ($B_\phi^2 / B_{\bf p}^2$) for our BNS merger simulations as a function of time for the remnant (top) and the disk (bottom).
In the remnant, before the merger, the poloidal component dominates for all simulations except \POLLR, where the initial dipolar field is aligned with the equatorial plane and gives rise to an oscillatory toroidal component when measured from the center of mass of the binary.
Immediately after the merger, the toroidal magnetic field energy starts to dominate over the poloidal field energy both in the disk and the remnant and remains dominant until the end of the simulations.
All simulations show slow secular growth in the ratio of these two quantities from the merger up to ${\sim} 20$ ms post-merger.
After this time, the ratio saturates or slightly decreases, indicating a larger relative growth of the poloidal component of the field while the toroidal field reaches a plateau.
The remnants show a large variance in the maximum value of this ratio among the simulations, with the purely toroidal runs reaching the maximum value at $\langle B_\phi^2 / B_{\bf p}^2 \rangle \sim 30$.
Simulation \POLQ\ consistently shows the lowest ratio for our simulation sample.
In the disk, the toroidal-to-poloidal magnetic energy ratio is also $>1$, but its evolution is more chaotic than that in the remnant; the toroidal field dominates by only a factor of ${\sim} 2{-}5$ at most.

Figure \ref{fig:avg_invbeta} shows the averaged magnetic-to-gas pressure ratio $\langle \beta^{-1} \rangle$. This evolves similarly to the average magnetic field intensity.
The magnetization in the proto-disk is very low before the merger, but it rapidly grows after it.
After an initially exponential growth, $\langle \beta^{-1} \rangle$ keeps secularly growing as $\sim t^2$.
In the remnant, the evolution is slightly different.
After the exponential amplification of the magnetic field (and thus of $\beta^{-1}$) due to the KH instability, most of the simulations show a saturation or at most a slow growth, while for \BITANT, the magnetization keeps growing at an exponential rate, consistent with the magnetic field evolution.

The evolution of the average entropy per baryon in the disk and the remnant is shown in Fig. \ref{fig:avg_s}.
This quantity evolves similarly across the various simulations, provided the EOS and the mass ratio are the same.
We have averaged the various curves for all the simulations that share these properties and compared them with the \STIFF, the \NOB, and the \POLQ\ simulation.
For the equal-mass merger simulations with SFHo, the entropy per baryon rapidly grows at the merger due to the compression produced in the collision of the NSs and then saturates at $s/k_{\rm B}\sim 12$ in the disk and $s/k_{\rm B}\sim 1.5$ in the remnant.
Magnetic fields seem to have a negligible impact on the bulk entropy evolution: the non-magnetized (\NOB) run closely matches the mass-weighted specific entropy of the magnetic runs.

Compared with the equal-mass SFHo cases, both \POLQ\ and \STIFF\ exhibit systematically lower mass-weighted entropy in the disk ($\Delta s/k_{\rm B} \sim 2$) and marginally higher values in the remnant ($\Delta s/k_{\rm B}\sim 0.2$).
The origin of the lower disk entropy differs between these two cases.
In \POLQ, the lower mass star is partially tidally disrupted before coalescence, depositing very-low entropy material ($s/k_{\rm B}< 1$) into the nascent disk.
In \STIFF, the stars (and the post-merger remnant) are less compact, so early post-merger remnant oscillations transfer a larger fraction of relatively low-entropy material ($s/k_{\rm B}\sim 1-2$) into the disk than in the SFHo cases.
The rapid addition of this low-entropy mass increases the mass of the disk (see Fig. \ref{fig:mass}) while reducing its mass-weighted specific entropy.
After this transient outflow subsides, continued shock and viscous heating drive the disk entropy back up.
The remnant's mean entropy is less affected because the transferred mass is small compared with the remnant's much larger mass and therefore has little impact on its mass-weighted average.
The evolution and spatial distribution of the entropy per baryon in our runs is broadly consistent with the results of Ref. \cite{Perego_etal2019EPJA...55..124P}, who additionally showed that neutrino emission/absorption does not significantly affect this quantity.

Finally, Fig. \ref{fig:mass} shows the total mass of the disk and remnant as a function of time for our simulations.
Again, we have averaged the mass evolution for all simulations with equal-mass binaries and the SFHo EOS, and compared it with the simulations with low or null magnetic field, a stiff EOS, or a larger mass ratio.
We also averaged the two HR runs, which show no differences from the equivalent SR runs.

All the simulations show a rapid increase in the mass of the disk at the merger, followed by a much slower but steady growth until the end of the simulation.
Since the disk mass comes from the remnant, the mass of the latter decreases consistently with time.
Naturally, \POLQ\ shows a larger disk mass, which starts to build up before the merger due to the tidal disruption of the lighter NS.
Simulation \STIFF\ also shows a larger disk mass and a lower remnant mass.
The stiff EOS simulation produces a larger remnant with less bound material in its outer boundaries, favoring an increase in the disk mass.
Interestingly, simulations with a lower or no magnetic field show a larger disk mass.
However, there is no clear trend since \LOWB\ shows a larger mass than \NOB\ simulation, which may indicate that this difference is due to numerical artifacts and not a real physical process.
On the other hand, the \BITANT\ run shows a slightly larger disk mass than average; however, the difference is within the uncertainties and not significant.
The measured values for the disk masses are consistent with those found by other authors for similar merger conditions (e.g., Refs. \cite{Prakash_etal2021PhRvD.104h3029P, Bernuzzi_etal2020MNRAS.497.1488B}).

\subsubsection{Radial and vertical distribution}
\label{sec:radial}

\begin{figure}
    \centering
    \includegraphics[width=0.999\linewidth]{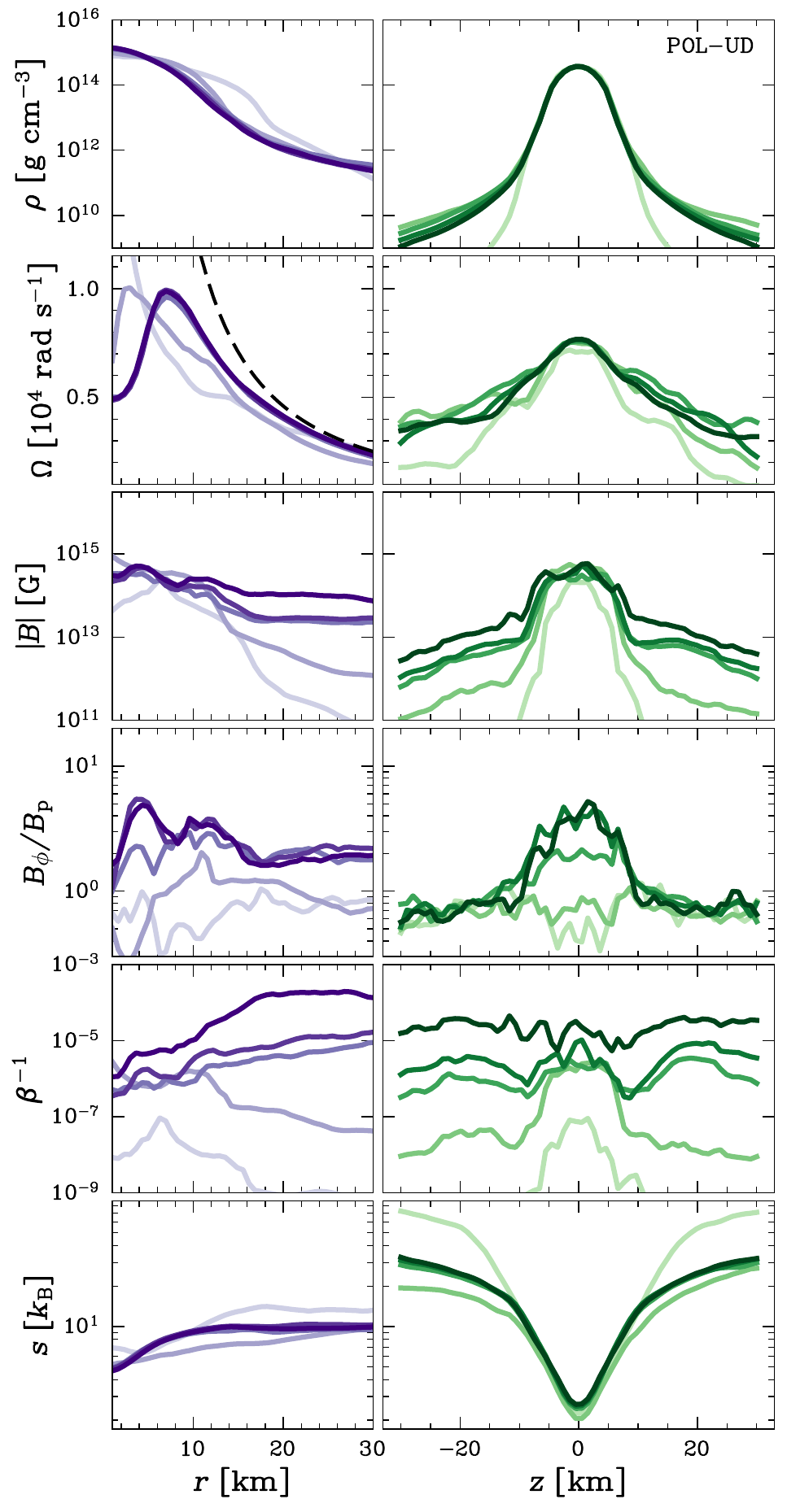}
    \caption{Radial (left column) and vertical (right column) profiles of several quantities for simulation \POLUD. From top to bottom: rest-mass density $\rho$; angular velocity $\Omega$; absolute value of the magnetic field $|B|$; toroidal-to-poloidal magnetic field ratio $B_\phi / B_{\rm p}$; magnetic-to-gas pressure ratio ($\beta^{-1}$).
    Each curve shows radial profiles for a single time of $t-t_{\rm mer}=-1,~ 1,~ 10,~20,~{\rm or}~28.3~{\rm ms}$, where darker colors correspond to later times.
    The dark dashed curve in the angular velocity radial profile shows a Keplerian profile, $\Omega \sim r^{-3/2}$.}
    \label{fig:profiles_fiducial}
\end{figure}

\begin{figure}
    \centering
    \includegraphics[width=0.999\linewidth]{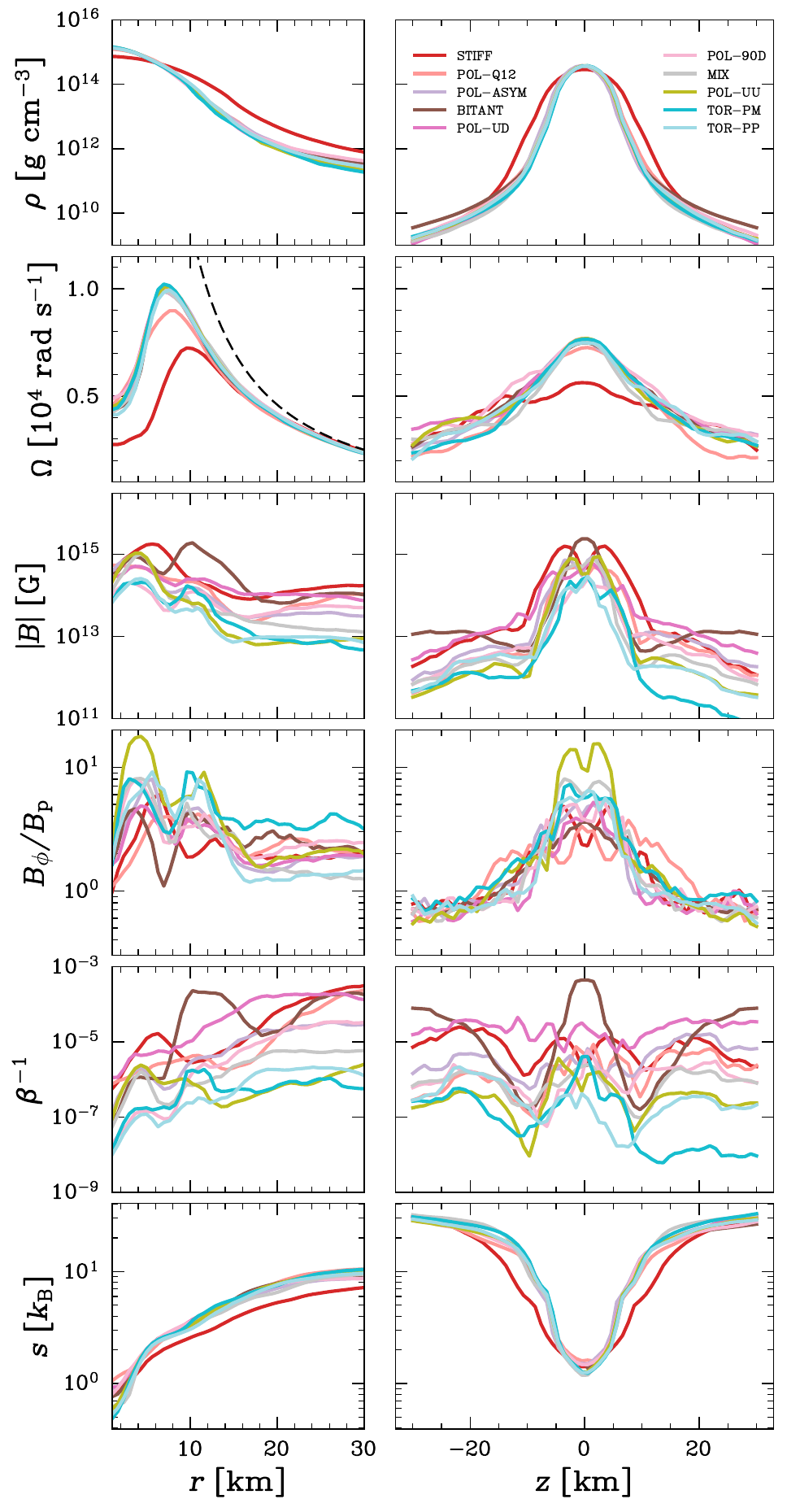}
    \caption{Radial (left column) and vertical (right column) profiles of several quantities for the various simulations at the last snapshot.
    The dark dashed curve in the angular velocity radial profile shows a Keplerian profile, $\Omega \sim r^{-3/2}$.}
    \label{fig:profiles_all}
\end{figure}

In this Section, we analyze the radial and vertical distribution of the magnetic field and other properties of the post-merger remnant and disk.
To compute radial profiles of magnetic field and thermodynamic quantities, we perform cylindrical averages at different radii, considering only material with $|z| < 15~{\rm km}$
On the contrary, for the rest-mass density and angular velocity profiles, we focus on $\phi$-averages on the equatorial plane at each radius.
Vertical profiles of all quantities are calculated by averaging on $z={\rm const.}$ surfaces restricted to regions close to the $z$-axis: $x^2+y^2 < 15~{\rm km}$.

Radial (vertical) profiles for simulation \POLUD\ are shown in the left (right) column of Fig. \ref{fig:profiles_fiducial}.
From top to bottom, we show the rest-mass density $\rho$, angular velocity $\Omega$, absolute value of the magnetic field $|B|$, toroidal-to-poloidal magnetic field ratio $B_\phi/B_{\bf p}$, magnetic-to-gas pressure ratio $\beta^{-1}$, and entropy per baryon $s$.
In each plot, the darker the line color, the later the time, starting from ${\sim} 1~{\rm ms}$ before the merger until the end of the simulation ${\sim} 30 ~{\rm ms}$.
The density profile on the equatorial plane initially shows a very oblate post-merger remnant, but it quickly settles to a more stable shape as the outer layers are ejected on the equatorial plane.
Similarly, the angular velocity rapidly settles to a differentially rotating profile with $d\Omega / dR > 0$ within the core of the remnant, reaching a maximum angular velocity of ${\sim} 10~{\rm rad~ms}^{-1}$ at $R \sim 7~{\rm km}$; this corresponds to a rotational period of ${\sim} 0.6~{\rm ms}$.
At larger radii, the profile slowly transitions to a quasi-Keplerian disk ($\Omega \sim r^{-3/2}$), whose angular velocity profile is overplotted with a black dotted curve.

The magnetic field is initially largely confined within the remnant.
Immediately after the merger, the initial KH amplification phase increases the absolute value of $B$ within the remnant.
At the same time, the disk is populated with material with a high magnetization that is expelled from the remnant.
In most of our simulations, the quality factor for MRI throughout the disk is $Q_{\rm MRI}=\lambda_{\rm MRI}/\Delta x \lesssim 10$, where $\lambda_{\rm MRI} = B_{\bf p}/\sqrt{4\pi \rho c^2} \times (2\pi / \Omega)$ is the wavelength of the fastest axisymmetric growing mode; hence, it cannot be guaranteed that the MRI is well resolved but slower growing modes could still be playing a role in the amplification of the disk magnetic field.
At more evolved times ($t-t_{\rm mer} \gtrsim 20~{\rm ms}$), the magnetic file profile becomes approximately constant in radius, with values ${\sim} 10^{14}~{\rm G}$ up to $r=30~{\rm km}$, consistently with the saturation we described in Sec. \ref{sec:thermo}.
The vertical profile also shows a steady increase in the magnetic field at all heights with no convergence in time.

The toroidal-to-poloidal magnetic field ratio shows a different type of evolution in the radial and vertical directions.
The radial profile shows that the initially poloidal field rapidly becomes subdominant to the generated toroidal field after the merger within the remnant and the disk. This ratio seems to stabilize to a rather constant value of ${\sim} 2{-}3$ in the outer parts of the remnant and the disk, and peaks in the remnant core at a value ${\sim} 4{-}5$.
In the vertical direction, the toroidal component increases its dominance with time within the remnant but the poloidal field dominates for $z \gtrsim 10~{\rm km}$, reaching a ratio ${\sim} 0.5{-}0.7$ which does not evolve significantly at late times; this indicates that both poloidal and toroidal components are growing at comparable rates in the polar funnel.

The above considerations are consistent with the evolution of the magnetic-to-gas pressure ratio.
The relative importance of the magnetic field over the gas pressure keeps growing with time, both radially and vertically, and does not saturate at the latest times.
In the left plot, it can be seen that $\beta^{-1}$ grows faster at larger radii and lower in the remnant.
This follows the population of these regions with high magnetic field material that escaped the remnant, as well as in-situ amplification of the magnetic field.
In the final state, $\beta^{-1}\lesssim 10^{-5}$ within the remnant and grows to $\beta^{-1}\sim 10^{-4}$ towards the central parts of the disk.
In the vertical direction, the magnetic-to-gas pressure ratio grows very rapidly outside the remnant and slower within it, reaching a constant value in height of $\beta^{-1}\sim 10^{-5}$ in the final state.

Finally, the entropy per baryon marginally evolves during the post-merger phase.
It rapidly saturates to an approximately constant value of $s/k_{\rm B} \lesssim 10$ in the radial direction, but it grows in the vertical direction from a minimum of $s/k_{\rm B}\sim 2$ at $z=0$ to $s/k_{\rm B}\sim 30$ at $z\sim 30~{\rm km}$.

Figure \ref{fig:profiles_all} shows radial and vertical profiles of the above-mentioned quantities at the last time of each of our simulations.
As expected, the rest-mass density is not affected by the different magnetic field configurations, but only by the EOS: simulation \STIFF\ produces a less dense and more extended remnant and a denser disk than simulations with the softer SFHo EOS.
The EOS also affects the rotational profile for the \STIFF\ simulation, which reaches a peak at a larger radius of ${\sim} 9{-}10~{\rm km}$ and a longer maximum rotational period of ${\sim} 0.9 ~{\rm ms}$.
Simulation \POLQ\ also shows a slightly different rotational profile within the remnant with a lower maximum angular velocity than the equal-mass binaries.
In the vertical direction, the trends are similar, with only \STIFF\ producing a different profile from the others due to the more extended remnant.

The magnetic field profiles present larger differences among the simulations, consistent with the findings of the previous section.
Most of the simulations show $d\beta^{-1} / dR > 0$ as we described above.
Simulation \POLLR\ shows the least magnetized remnant core, but a value of the magnetic field in the disk consistent with the other runs.
Since this run also shows the largest $B_\phi / B_{\rm p}$ value within the remnant, the low magnetic field corresponds to a much lower poloidal field than the rest of the simulations, which is a consequence of its low initial value.
Finally, simulation \BITANT\ shows the largest magnetic field and magnetization in the outer regions of the remnant, despite its last snapshot corresponding to a time ${\sim} 15{-}20~{\rm ms}$ earlier than the rest of the simulations.

The vertical profiles of magnetic field quantities also show large differences among the simulations.
Simulation \BITANT\ shows the largest magnetic field and magnetization on the equatorial plane due to a large poloidal magnetic field forced by the boundary conditions.
It shows a sharp drop at $|z| \sim 10 ~{\rm km}$, and an increase again above that height to reach the largest value at $|z| \sim 40~{\rm km}$.
The other simulations reach a more constant $\beta^{-1}(z)$ profile but with different final average values in the range $\beta^{-1} \sim 10^{-7}-3\times 10^{-5}$.
Simulation \POLUU, in a similar way to the behavior in the radial profile, shows the lowest magnetization in the vertical direction, with $\beta^{-1} \sim 3 \times 10^{-8}$ for $|z| > 10~{\rm km}$.

Finally, the entropy per baryon again is largely independent of the initial magnetic field setup and only varies significantly with the EOS.
Simulation \STIFF\ shows the lowest overall value of $s/k_{\rm B}$ both in the radial and vertical directions.

\section{Discussion}
\label{sec:discussion}

The evolution of the magnetic field in our simulations follows distinct patterns in different regions of the post-merger system.
Before the merger, the magnetic field is determined by the initial conditions, for which it is predominantly coherent and it is confined within the neutron stars (see Fig. \ref{fig:fiducial_xy}).
Following the merger, it first suffers an exponential amplification driven by the KH instability at the shear interface that forms between the two colliding NSs.
The initially coherent field rapidly amplifies and becomes largely turbulent after the merger; at later times, however, a coherent toroidal field slowly builds up again within the remnant (see Fig. \ref{fig:fiducial_xz}).
This part of the evolution is dominated by magnetic winding caused by the differential rotation of the remnant.

The toroidal field dominates over the poloidal field independently of the setup and this persists during the simulations (see Figs. \ref{fig:bphi2_bp2} and the 4th row of Fig. \ref{fig:profiles_all}), with the toroidal-to-poloidal energy ratio stabilizing around $2{-}4$ in most cases within the disk but with a much larger deviation (${\sim} 8-30$) within the remnant.
This ratio is particularly important for understanding the potential launching of relativistic jets, as the field configuration strongly influences the efficiency of energy extraction mechanisms.
The magnetic-to-gas pressure ratio $\beta^{-1}$ shows a secular growth in our simulations, particularly in the disk (Fig. \ref{fig:avg_invbeta}).
It does not reach saturation within the simulations' lifespan, suggesting that longer-term evolution might lead to even higher magnetization, potentially reaching the values necessary for jet launching at later times.
The radial profiles shown in Figs. \ref{fig:profiles_fiducial} and \ref{fig:profiles_all} reveal that $\beta^{-1}$ grows faster with time at larger radii, indicating that the disk becomes increasingly magnetically dominated over time.
The growing magnetization in the disk may provide favorable conditions for launching outflows, though they are not sufficient yet to show a clear difference in the amount of ejected mass (see Paper II).
The strong toroidal field in the remnant could support the formation of a central engine for sGRBs. 

As shown in Fig. \ref{fig:fiducial_xz_all}, simulations with initially vertical and aligned dipolar fields develop toroidal fields with a counter-clockwise (clockwise) direction above (below) the equatorial plane.
In contrast, simulations with initially anti-aligned poloidal field (\POLUD) or a toroidal field (\TORPP) give rise to toroidal fields which exhibit the opposite polarity.
Despite these differences in field topology, the overall energetics of the system --particularly the toroidal-to-poloidal energy ratio and the $\beta^{-1}$ parameter-- show consistency across different initial configurations.
Interestingly, the simulation with bitant symmetry (\BITANT) shows a significantly more coherent field structure on the equatorial plane even at early times after the merger, indicating that the imposed symmetry condition reduces the development of turbulence.
This has methodological implications for future simulations, as symmetry assumptions may artificially suppress certain aspects of the magnetic field evolution.

Regarding the dependence of the post-merger evolution with the EOS, \STIFF\ simulation produces a less dense, more extended remnant with a slower rotation rate, reaching a maximum angular velocity at a larger radius (${\sim} 9{-}10 ~ {\rm km}$) compared to the SFHo EOS.
Nevertheless, the larger volume of the remnant and the weaker collision help the initial poloidal field to survive after the merger; hence, the toroidal field generated by winding seems to be affected by the initially large poloidal field.
The thermodynamic properties also differ between the two EOS models, with the \STIFF{ simulation showing lower entropy per baryon both radially and vertically.
This has implications for the thermal pressure support in the remnant and disk, potentially affecting the long-term stability of the system and the conditions for jet formation.
Future gravitational wave observations that constrain the neutron star equation of state could then provide information about the expected magnetic field configurations.

A similar study to ours was performed in Ref. \cite{Kawamura_etal2016PhRvD..94f4012K}.
They investigated the influence of the pre-merger magnetic field conditions, the binary mass ratio, or the nuclear EOS on the post-merger evolution for BNS mergers that produce an early collapse to a BH at $\sim 10$ ms after the merger.
They only consider initial poloidal magnetic fields with a maximum value of $B_{\rm max}=10^{12}~{\rm G}$ and with different relative orientations of the field for each star and assumed either an ideal gas EOS or the H4 EOS; their runs also assume reflection symmetry across the equatorial plane.
Our results and theirs can be compared only during the first ${\sim} 10~{\rm ms}$ after the merger, before BH formation.
Opposite to them, we found the largest amplification in the early post-merger phase for the antialigned poloidal field case, though it rapidly decays later to values matching those in the other scenarios.

Ref. \cite{AguileraMiret_etal2022ApJ...926L..31A} found that the pre-merger magnetic field topology is quickly forgotten after the merger, and the final magnetic field intensity and structure are similar independently of the initial setup.
Their simulations made use of a subgrid turbulence modelling scheme, which allows them to reach large post-merger magnetic fields from initial realistic magnetic field values ($B<10^{12}~{\rm G}$).
Thus, their conclusions are valid if the pre-merger magnetic fields are small and they depend on the assumed parameters for their subgrid modeling.
In our case, we have shown that, at least for artificially large magnetic fields commonly assumed in the literature, the initial conditions do have an effect on the post-merger evolution.

\section{Conclusions and future directions}
\label{sec:conclusions}

We have presented an analysis of a set of GRMHD simulations of binary neutron star mergers run with the code \texttt{GR-Athena++} to study the post-merger evolution of magnetic fields in these events.
We explored the differences among scenarios that differ either in the initial magnetic field topology, the nuclear EOS, or the binary mass ratio.
We also compared our runs with simulations where we assumed reflection symmetry across the equatorial plane, and with simulations with a very low or a zero magnetic field, and with high-resolution simulations.

The initial topology of the magnetic field impacts its amplification, predominantly within the remnant.
In addition, imposing reflection symmetry across the equatorial plane produced two noticeable effects: {\it i)} a decrease in the strength of the turbulence developed close to the equatorial plane compared with the other runs, and {\it ii)} a faster and stronger amplification of the magnetic field.
These two effects are likely a consequence of the assumed boundary condition, which forces the magnetic field to be poloidal at the equatorial plane.
It is worth mentioning that some of the largest simulations published to date assume this symmetry to reduce computational costs.

In the post-merger phase, magnetic fields are predominantly toroidal in both the remnant and the surrounding disk, though not in the polar funnel, where the poloidal magnetic field dominates.
The magnetic field intensity in the disk suffers a rapid growth during the merger and then stabilizes or keeps growing at a slow rate.
The magnetization keeps growing until the end of the simulations, suggesting that at later times, magnetically dominated regions may form, possibly resulting in the launching of magnetized outflows.

In all the simulations, the mass of the disk rapidly grows at the merger and then saturates or grows at a low rate.
Simulation \POLQ\ achieves the largest disk mass, ${\sim} 0.14~M_\odot$, due to the tidal disruption of the lighter star before the merger, followed by \STIFF, which, due to the larger volume of the remnant, allows more material to be loosely bound to the remnant and populate the disk.
The rest of the simulations show consistent disk mass values of $\lesssim 0.1~M_\odot$.
The influence of the magnetic field on the disk mass is unclear, since whereas \LOWB\ shows a larger mass than the other runs, simulation \NOB\ behaves similarly to the high-$B$ runs.

In future studies, we will explore longer post-merger timescales to determine the conditions for the formation of relativistic jets; we will also include neutrino transport, providing a more complete picture of the thermal evolution of the system and its impact on mass ejection and nucleosynthesis.
Finally, we will explore a wider range of initial magnetic field strengths, beyond the $B = 5\times 10^{15} {\rm G}$ used in this study, to determine the dependence of the post-merger evolution on the initial magnetization of the neutron stars.

\begin{acknowledgments}
  We thank the referee for the useful comments.
  EG thanks Luciano Combi for helpful discussions.
  EG and DR acknowledge funding from the National Science Foundation under Grants No.~AST-2108467 and PHY-2407681.
  EG acknowledges funding from an Institute for Gravitation and Cosmology fellowship.
  JF and DR acknowledge U.S. Department of Energy, Office of Science, Division of Nuclear Physics under Award Number(s) DE-SC0021177 and DE-SC0024388.
  BD, SB, and MJ acknowledge support by the EU Horizon under ERC Consolidator Grant,
  no. InspiReM-101043372.
  PH acknowledges funding from the National Science Foundation under Grant No. PHY-2116686.
  The authors are indebted to A.~Celentano's PRİSENCÓLİNENSİNÁİNCIÚSOL.
  The numerical simulations were performed on TACC's Frontera (NSF LRAC
  allocation PHY23001) and on Perlmutter. This research used resources
  of the National Energy Research Scientific Computing Center, a DOE
  Office of Science User Facility supported by the Office of Science of
  the U.S.~Department of Energy under Contract No.~DE-AC02-05CH11231.
  Simulations were also performed on the national HPE Apollo Hawk
  at the High Performance Computing Center Stuttgart (HLRS).
  The authors acknowledge HLRS for funding this project by providing access
  to the supercomputer HPE Apollo Hawk under the grant numbers INTRHYGUE/44215
  and MAGNETIST/44288.
  Simulations were also performed on SuperMUC\_NG at the
  Leibniz-Rechenzentrum (LRZ) Munich.
  The authors acknowledge the Gauss Centre for Supercomputing
  e.V. (\url{www.gauss-centre.eu}) for funding this project by providing
  computing time on the GCS Supercomputer SuperMUC-NG at LRZ
  (allocations {\tt pn67xo}, {\tt pn76li}, {\tt pn68wi} and {\tt pn36jo}).
  Postprocessing and development run were performed on the ARA cluster
  at Friedrich Schiller University Jena.
  The ARA cluster is funded in part by DFG grants INST
  275/334-1 FUGG and INST 275/363-1 FUGG, and ERC Starting Grant, grant
  agreement no. BinGraSp-714626.
\end{acknowledgments}

\bibliographystyle{unsrtnat}
\bibliography{refs}





\end{document}